\documentclass[letterpaper]{article} % DO NOT CHANGE THIS
\usepackage{aaai2026}  % DO NOT CHANGE THIS
\usepackage{times}  % DO NOT CHANGE THIS
\usepackage{helvet}  % DO NOT CHANGE THIS
\usepackage{courier}  % DO NOT CHANGE THIS
\usepackage[hyphens]{url}  % DO NOT CHANGE THIS
\usepackage{graphicx} % DO NOT CHANGE THIS
\usepackage{natbib}  % DO NOT CHANGE THIS AND DO NOT ADD ANY OPTIONS TO IT
\usepackage{caption} % DO NOT CHANGE THIS AND DO NOT ADD ANY OPTIONS TO IT
\usepackage{algorithm}
\usepackage{algorithmic}

\usepackage{subcaption}
\usepackage{booktabs}
\usepackage{amsmath}
\usepackage{multirow}
\usepackage{xcolor}
\usepackage{etoolbox}
\usepackage{enumitem}
\usepackage{etoolbox}
\usepackage{environ}
\usepackage{xparse}
\usepackage{ragged2e} 
\usepackage[bb=boondox,bbscaled=.95,cal=boondoxo]{mathalfa} 
\newcommand{\blockquoteauthor}{} % temporary storage
{%
  \renewcommand{\blockquoteauthor}{#1}%
  \begin{list}{}{\leftmargin=1em \rightmargin=1em}
  \item\relax
  \small
  \setlength{\parindent}{1em}
  \justifying
}%
{%
  \end{list}
  \ifstrempty{\blockquoteauthor}{}{%
    \vspace{-0.5em}
    \begin{flushright}
      \small\itshape-- \blockquoteauthor
    \end{flushright}
  }%
  \smallskip
}

\usepackage{newfloat}
\usepackage{listings}
\DeclareCaptionStyle{ruled}{labelfont=normalfont,labelsep=colon,strut=off} % DO NOT CHANGE THIS
\floatstyle{ruled}
\newfloat{listing}{tb}{lst}{}
\floatname{listing}{Listing}
\nocopyright 
\title{Information Pathways in Online Science Communication: The Role of Platform Actors and News Media}
\author {
    Alexandros Efstratiou\textsuperscript{\rm 1},
    Giuseppe Russo\textsuperscript{\rm 2},
    Luca Luceri\textsuperscript{\rm 3}
}
\affiliations {
    \textsuperscript{\rm 1}University of Washington\\
    \textsuperscript{\rm 2}EPFL\\
    \textsuperscript{\rm 3}Information Sciences Institute, University of Southern California\\
    aefstra@uw.edu, giuseppe.russo@epfl.ch, lluceri@isi.edu
}
\usepackage{bibentry}
\begin{document}

\maketitle

\begin{abstract}
Online discussions of science involve complex interactions among experts, news media, and social media users as they interpret and disseminate scientific findings. While prior work has examined these actors in isolation, their interplay in shaping science communication remains poorly understood. Using the COVID-19 pandemic as a case study, we analyze 1.24M tweets and 211k news articles that reference pandemic-related scientific papers. We find that the most influential Twitter accounts in this discourse are predominantly individuals with medical or research credentials. However, we also identify a coordinated network that disproportionately amplifies a small set of prominent credentialed experts who advance contrarian, anti-consensus positions on vaccines, lockdowns, and related topics. The papers promoted by these influential actors substantially overlap with those covered by news media, but with key differences: pro-consensus experts primarily engage with studies featured by mainstream and medical outlets, whereas contrarian experts align more closely with papers promoted by low-quality, pseudoscientific, or conspiratorial sources. Notably, news outlets tend to report on scientific studies after they have been highlighted by social media superspreaders. Together, these findings reveal multi-level pathways of information flow and coordinated amplification structures that shape science communication across social media and news, offering new insights into the dynamics of the broader information ecosystem.
% Online discussions of science often involve experts, news media, and social media users trying to make sense of scientific information.
% While the roles of these groups have been studied individually, their interplay remains less understood in the context of science communication. 
% Using COVID-19 as a case study, we analyze a dataset of 1.24M tweets and 211k news articles discussing pandemic-related scientific papers.
% We find that most influential Twitter accounts in this discourse are people with medical or research credentials. 
% However, we detect a coordinated network that amplifies posts by a minority of prominent contrarian credentialed experts who hold anti-consensus views on vaccines, lockdowns, and related topics. 
% The scientific papers promoted by these influential accounts overlap with those highlighted in news media: pro-consensus experts tend to discuss papers featured by mainstream or medical news outlets, whereas contrarian experts align more with low-quality, pseudoscientific, or conspiratorial outlets.
% Notably, news outlets typically report on scientific studies after superspreaders mention them.
% This paper highlights multi-level pathways of information flow and the coordination structures shaping science communication across social media and news, paving the way for future research into the broader information ecosystem.

\end{abstract}


\section{Introduction}\label{sec:intro}

% Skepticism around scientific facts that have long been upheld by empirical evidence, like climate change~\cite{hornsey_toolkit_2022} and the benefits of vaccination~\cite{paoletti_political_2024}, has been on the rise over the past decade~\cite{kennedy_americans_2023}, driven primarily by increasingly polarized attitudes towards science~\cite{milkoreit_rapidly_2025}.
% Although much of this skepticism may arise from politically motivated reasoning or disinformation campaigns~\cite{lewandowsky_liars_2024}, recent evidence suggests that, for scientific topics, even misinformed people pay attention to scientific credentials~\cite{jalbert_who_2025}, and those with perceived scientific credentials may come to dominate even anti-science online communities~\cite{efstratiou_misrepresenting_2021,harris_perceived_2024}.
Skepticism toward scientific facts long supported by empirical evidence—such as climate change~\cite{hornsey_toolkit_2022} and the benefits of vaccination~\cite{paoletti_political_2024}—has increased over the past decade~\cite{kennedy_americans_2023}, driven in part by growing polarization in public attitudes toward science~\cite{milkoreit_rapidly_2025}. While this skepticism is often attributed to politically motivated reasoning or organized disinformation campaigns~\cite{lewandowsky_liars_2024}, recent research suggests a more nuanced dynamic: even individuals who hold misinformed views on scientific issues attend to scientific credentials~\cite{jalbert_who_2025}, and actors perceived as scientifically authoritative can come to dominate discourse within anti-science online communities~\cite{efstratiou_misrepresenting_2021,harris_perceived_2024}.

% Meanwhile, science dissemination and communication can also take a grassroots form online.
% Several works have documented how scientific papers are shared among online communities for reasons ranging from offering collaborative pushback to scientific findings~\cite{yudhoatmojo_we_2021}, to selectively misrepresenting scientific consensus~\cite{beers_selective_2023, efstratiou_heres_2024,yudhoatmojo_understanding_2023}, or occasionally, even for scientists to hold open, scientific discussions in public forums~\cite{efstratiou_heres_2024}.
% Beyond that, scientific misinformation can also spread in the form of sub-standard community data analysis efforts~\cite{lee_viral_2021} or pseudoscientific content~\cite{papadamou_it_2022}.
At the same time, science dissemination and interpretation increasingly occur through grassroots activity in online spaces. Prior work shows that scientific papers are shared within online communities for a range of purposes, including coordinated challenges to established findings~\cite{yudhoatmojo_we_2021}, selective misrepresentation of scientific consensus~\cite{beers_selective_2023,efstratiou_heres_2024,yudhoatmojo_understanding_2023}, and open scientific debate conducted by researchers in public forums~\cite{efstratiou_heres_2024}. Beyond the circulation of peer-reviewed research, scientific misinformation also emerges through substandard community-driven data analyses~\cite{lee_viral_2021} and the production or amplification of pseudoscientific content~\cite{papadamou_it_2022}.

% Although the centralized influence of perceived experts~\cite{harris_perceived_2024}, some of whom occasionally have vested interests in promoting alternative narratives~\cite{lewandowsky_liars_2024,nogara_disinformation_2022} is well-documented, less is understood about how such grassroots, distributed efforts may cross over into centralized expert influence or even outright malicious disinformation.
% Moreover, entities like news media occasionally play a substantial role in science dissemination towards wider audiences~\cite{west_misinformation_2021}, but less is understood about their role in overarching information pathways.

Although the centralized influence of perceived experts is well documented~\cite{harris_perceived_2024}, including cases where such actors have vested interests in promoting alternative narratives~\cite{lewandowsky_liars_2024,nogara_disinformation_2022}, much less is known about how grassroots, distributed efforts evolve into centralized expert influence or even into coordinated, malicious disinformation. Moreover, while news media can play a substantial role in disseminating scientific information to broader audiences~\cite{west_misinformation_2021}, their position within the overarching pathways of science-related information flow remains poorly understood.

\paragraph{Research questions.}

% In this paper, we use COVID-19 science as a case study to offer a more holistic look into a wider science dissemination ecosystem.
% We focus on information pathways between different Twitter actors, namely, users, bots, superspreaders, and coordinated accounts, as well as news media articles, to characterize their distinct activities but also to examine how their actions are intertwined.
In this paper, we use COVID-19 science as a case study to examine the science dissemination ecosystem from a holistic perspective. We analyze \textit{information pathways} linking different classes of Twitter actors, namely, organic users, bots, superspreaders, and coordinated accounts, as well as news media articles, characterizing their distinct roles while examining how their activities intersect and reinforce one another.
% We define information pathways as the order of precedence between different types of entities taking similar actions; for example, influential users and news media discussing the same scientific pieces, or users retweeting other users' science-related posts.
% We stress that information pathways under this definition are not necessarily causal.
% For example, news outlets reporting on a paper after influential Twitter users brought it into public attention would mean that Twitter \textit{preceded} the news in this case, but the news reporting could also have occurred due to several other factors.
We define information pathways as patterns of temporal precedence between different types of entities engaging in similar actions, such as influential users and news media discussing the same scientific papers, or users retweeting science-related posts from other users. Importantly, information pathways under this definition are not inherently causal. For instance, if news outlets report on a scientific paper after it has been highlighted by influential Twitter users, Twitter is said to precede the news in this pathway, even though the news coverage may have emerged independently or in response to other external factors.

% We utilize a dataset of COVID-related preprints' mentions across Twitter and news media to answer the following research questions:
To examine how scientific information circulates across social media and news media, we leverage a dataset capturing mentions of COVID-19–related preprints on Twitter and in news articles. Using this data, we address the following research questions:

% \begin{itemize}[align=left]
%     \item[\textbf{RQ1.}] Who are the key actors responsible for disseminating COVID-19-related scientific information on Twitter, and how do they contribute to the amplification of research papers?
%     \item[\textbf{RQ2.}] Is there evidence of information pathways between Twitter and news media? Specifically, is research discussed by certain groups on Twitter subsequently picked up by media outlets, or vice versa?
% \end{itemize}

\begin{itemize}[align=left]
\item[\textbf{RQ1.}] How do different classes of Twitter actors contribute to the amplification and circulation of COVID-19-related scientific papers?
\item[\textbf{RQ2.}] What information pathways characterize the flow of COVID-19 scientific research between Twitter actors and news media?
\end{itemize}

\paragraph{Methods.}

% To address these questions, we employ a range of methods to detect and characterize the actors involved in online science communication. 
% We use co-activity networks to identify coordinated users, bot detectors to flag automated accounts, and activity metrics to measure influence. 
% We then characterize these users by their popularity, topical focus, emotional tone, and alignment with scientific consensus. 
% For this paper, we define \textit{contrarians} as users who typically hold anti-consensus views (e.g., opposition to COVID vaccines), and \textit{conformists} as users who align with scientific consensus. 
% In addition, we analyze potential information pathways from \textit{superspreaders} -- the most influential users -- and from news media, tracing how academic research papers diffuse within and beyond the Twitter information landscape.
To address these questions, we employ a suite of methods to identify and characterize actors involved in online science communication. We construct co-activity networks to detect coordinated users, apply bot-detection tools to flag automated accounts, and use activity-based metrics to quantify influence. We then profile these actors in terms of popularity, topical focus, emotional tone, and alignment with scientific consensus. Throughout the paper, we define \textit{contrarians} as users who consistently express anti-consensus positions (e.g., opposition to COVID-19 vaccines), and \textit{conformists} as users whose views align with established scientific consensus. Finally, we analyze potential information pathways originating from superspreaders---the most influential users---and from news media, tracing how academic research papers diffuse within and beyond the Twitter information landscape.

% To address these questions, we deploy a range of methods to detect and characterize the different actors participating in online science communication.
% We utilize co-activity networks to detect coordinated users, as well as bot detectors and influential activity metrics to find automated and highly influential users, respectively.
% We characterize them based on features like how popular they are, what kind of topics or emotional tone they use in their communications, and whether they are ``contrarian'' or ``conformist''.
% For the purposes of this paper, contrarian refers to users who typically hold anti-consensus views (e.g., anti-COVID vaccine), while conformists are users who align with scientific consensus.
% Moreover, we analyze potential information pathways from ``superspreaders'', i.e., the most influential users, and news media, examining the diffusion of  academic research papers within and across the Twitter information landscape.

\paragraph{Main findings.}

% We detect a contrarian (anti-consensus) coordinated network of accounts that copiously retweet similar content within a narrow time window.
% The main superspreaders are primarily conformist users with scientific credentials, along with some contrarian users and experts.
% However, the coordinated network tends to predominantly amplify content generated by these contrarian superspreaders.
% Although we find no evidence of \textit{automated} coordination in the sharing of COVID-19 science, the activity of news media aligns with the activity of superspreaders.
% Whereas more mainstream and trustworthy outlets tend to discuss similar papers as the conformist superspreaders, low-trust and conspiratorial/pseudoscientific outlets discuss similar papers as the contrarian superspreaders.
% Importantly, news outlet reporting typically follows the activity of superspreaders on Twitter, indicating that COVID-19 science usually gains traction on Twitter before making it into the news.
% However, this pathway may not necessarily be causal.
% Overall, our findings highlight the complex interplay between different platform actors and news media in the context of science communication, revealing how information pathways between them can also obfuscate scientific consensus.
We identify a coordinated network of contrarian (anti-consensus) accounts that repeatedly retweet similar content within narrow time windows. While the primary superspreaders in the dataset are largely conformist users with scientific credentials---alongside a smaller number of contrarian experts---the coordinated network disproportionately amplifies content produced by contrarian superspreaders. We find no evidence that this coordination is automated.
Moreover, news media activity exhibits temporal alignment with the activity of superspreaders. Mainstream and higher-trust outlets tend to cover the same scientific papers as conformist superspreaders, whereas low-trust, conspiratorial, or pseudoscientific outlets align more closely with contrarian superspreaders. Importantly, news coverage typically follows superspreader activity on Twitter, suggesting that COVID-19 scientific research often gains visibility on social media before appearing in news reporting—though this temporal ordering should not be interpreted as causal. Taken together, these findings underscore the complex interplay between social media actors and news outlets in science communication, revealing how information pathways across platforms can both amplify and obscure scientific consensus.
% These low-trust outlets are more likely to report on papers after they have already been discussed by contrarian superspreaders.
% On the other hand, mainstream media tend to precede conformist superspreaders in their reporting of papers to a similar extent as conformist superspreaders precede mainstream media.

\section{Related Work}

In this section, we cover existing work on the online science communication ecosystem, as well as work on online coordination more broadly.

\subsection{Science (Mis)communication}

Science communication has taken on a participatory form, where both experts~\cite{williams_hci_2022} and laypeople~\cite{hafid_scitweets_2022,yudhoatmojo_understanding_2023} often share scientific articles with their audiences and peers in online social media.
Additionally, beyond academic or other research venues, science can also be covered in other centralized formats like news media~\cite{pei_modeling_2025} or even documentaries~\cite{naiman_beyond_2025} and podcasts~\cite{gu_when_2025}.
This creates a particularly complex science communication environment.
On the one hand, science communicators may need to adopt platform-specific tactics to attract as much engagement as possible~\cite{bagchi_effects_2025,pera_shifting_2024}.
On the other hand, scientific misinformation is rampant~\cite{west_misinformation_2021}, especially around polarizing topics like climate change~\cite{uyheng_mainstream_2021} or COVID-19 vaccines~\cite{harris_perceived_2024,nogara_misinformation_2024}.
This may spread beyond decentralized contexts.
For example, \citet{bhat_scholarly_2025} report that scientists may withdraw from engaging with news media due to potential misrepresentations of their work.
% as one of the reasons.

The problem is compounded by how this science communication is taken up online.
It is common for discourse around scientific topics like vaccines and climate change to split into polarized clusters, each representing two sides of a debate~\cite{santoro_analyzing_2023,schmidt_polarization_2018,uyheng_mainstream_2021,zollo_debunking_2017}, and these biased, collective behaviors may also reinforce counter-science groups like the anti-vaccination movement~\cite{kata_anti-vaccine_2012}.
Nonetheless, signals of scientific expertise appear to be universally valuable for forming or supporting science-related attitudes~\cite{jalbert_who_2025}, even among people who do not otherwise hold consensus positions.
Online, this may manifest as scientific credentials being used to justify existing scientific opinions, rather than for knowledge attainment around scientific topics.
Recent works have shown that the way in which science is disseminated through social networks can lead to false consensus effects~\cite{efstratiou_heres_2024}, where science is occasionally selectively cited in public posts by anti-science communities in efforts to claim scientific evidence in their favor~\cite{beers_selective_2023}.
Moreover, perceived experts who share scientific articles at higher rates---but are, nonetheless, contrarians---are central in anti-vaccination communities~\cite{harris_perceived_2024}.
Taken together, this body of work suggests science contagion mechanisms that place ``desired'' experts at the center of communities, which may be inclined to use them as sources for their beliefs, although this potential social organization structure has yet to be empirically explored. 

\subsection{Social Collaboration and Coordination}

Social organization around science communication can be viewed through the lens of \textit{collaborative sensemaking}, a social process through which people make sense of complex information~\cite{dailey_its_2015}.
This process is occasionally driven by focused attention on certain information ``spotlights''~\cite{zhou_spotlight_2023}, and coordinating the agents involved can steer the quality and breadth of information that they are exposed to~\cite{hong_coordinating_2018}.
% This process may rely on distributed or centralized sources~\cite{krafft_centralized_2017}, and may be driven by focused attention on certain information ``spotlights''~\cite{zhou_spotlight_2023}.
% Tooling to assist with collaborative sensemaking has focused on coordinating the agents involved in the process~\cite{hong_coordinating_2018,mahyar_supporting_2014}, as this coordination steers the quality and breadth of information that the process participants are exposed to.

The idea of coordination, however, is also used in works that aim to detect potentially malicious activity.
This is often done by considering signals of shared actions, like sharing the same URLs or news domains~\cite{cinus_exposing_2025}, sharing the same social media posts~\cite{nizzoli_coordinated_2021,pacheco_uncovering_2021}, or posting content whose text is highly similar between users~\cite{pacheco_unveiling_2020}.
Common link-sharing on Facebook has been used to study coordination around COVID-19 misinformation~\cite{gruzd_how_2022} and COVID-19 vaccines more specifically~\cite{song_spread_2025}.
Occasionally, the timing of this shared activity is also considered~\cite{tardelli_temporal_2024}, and recent work has successfully deployed a combination of such signals in what are called ``fused networks'' to improve the generalizability of detection methods across information operations~\cite{luceri_unmasking_2024}.

However, these signals can also sometimes be attributed to organic, not inauthentic coordination~\cite{cao_organic_2015}, in line with collaborative sensemaking.
In such cases, we may expect that participants are drawn to shared entities in a bottom-up, organic manner, as with cherry-picking academic work that aligns with existing beliefs~\cite{beers_selective_2023,efstratiou_heres_2024}, or gravitating towards central, perceived experts~\cite{harris_perceived_2024,jalbert_who_2025}.
To this latter end, other scholars have also focused on detecting ``superspreaders''; users who post frequently, and whose posts tend to gain a lot of traction, within specific communities or topics~\cite{deverna_identifying_2024}.

\subsection{Present Work}

This paper aims to synthesize these insights to assess information pathways around science communication, using COVID-19 as a case study.
We focus specifically on scientific paper sharing, and examine the distinct roles of coordinated users, influential accounts (or ``superspreaders''), automated activity, and the news media.
Moreover, we attempt to link the activities of these distinct entities to each other, offering a first look into ``multi-level'' information pathways in science communication.

\section{Dataset}

We use a dataset from~\citet{efstratiou_heres_2024}. 
It includes over 25k COVID-related preprints (and their published versions, where available) released on bioRxiv and medRxiv from the start of the pandemic until November 4, 2022, along with data on their mentions in news and social media. 
The dataset contains 1.24M tweets about these papers posted by 346k users, and 211k news articles published by 2.34k unique outlets.
% , up to December 2022. 
Moreover, it labels users as contrarians or conformists based on a graph convolution of the retweet network, validated against annotated tweet content of sampled users.
In addition, papers are annotated by topic by applying a BERTopic model to their abstracts. 

% We make use of the dataset from~\citet{efstratiou_heres_2024}.
% It contains information on over 25k COVID-related preprints (and their published versions, where applicable) released on bioRxiv and medRxiv between the start of the COVID-19 pandemic and up until November 4th, 2022, alongside information about their mentions in news media and social media.
% Of note for the present work, the dataset contains 1.24M tweets about these papers posted by 346k users, as well as 211k news articles posted by 2.34k unique outlets, up until December 2022.
% Twitter users in the dataset are labeled as being either contrarians or conformists, based on a graph convolution of the retweet network.
% Moreover, the papers are annotated based on their topic through a BERTopic model that analyzes their abstracts.

\section{Characterizing On-Platform Actors}\label{sec:coordination}

In this section, we examine different actors on Twitter.
% , namely, potentially coordinated retweet networks, superspreaders, and automated bots.
We begin with the detection of coordinated retweet networks, followed by bots and superspreaders.
Finally, we consider the interaction between the coordinated retweet network and the superspreaders.

\subsection{Detection of Coordinated Retweet Networks}
\begin{figure}
    \centering
    \includegraphics[width=0.45\textwidth]{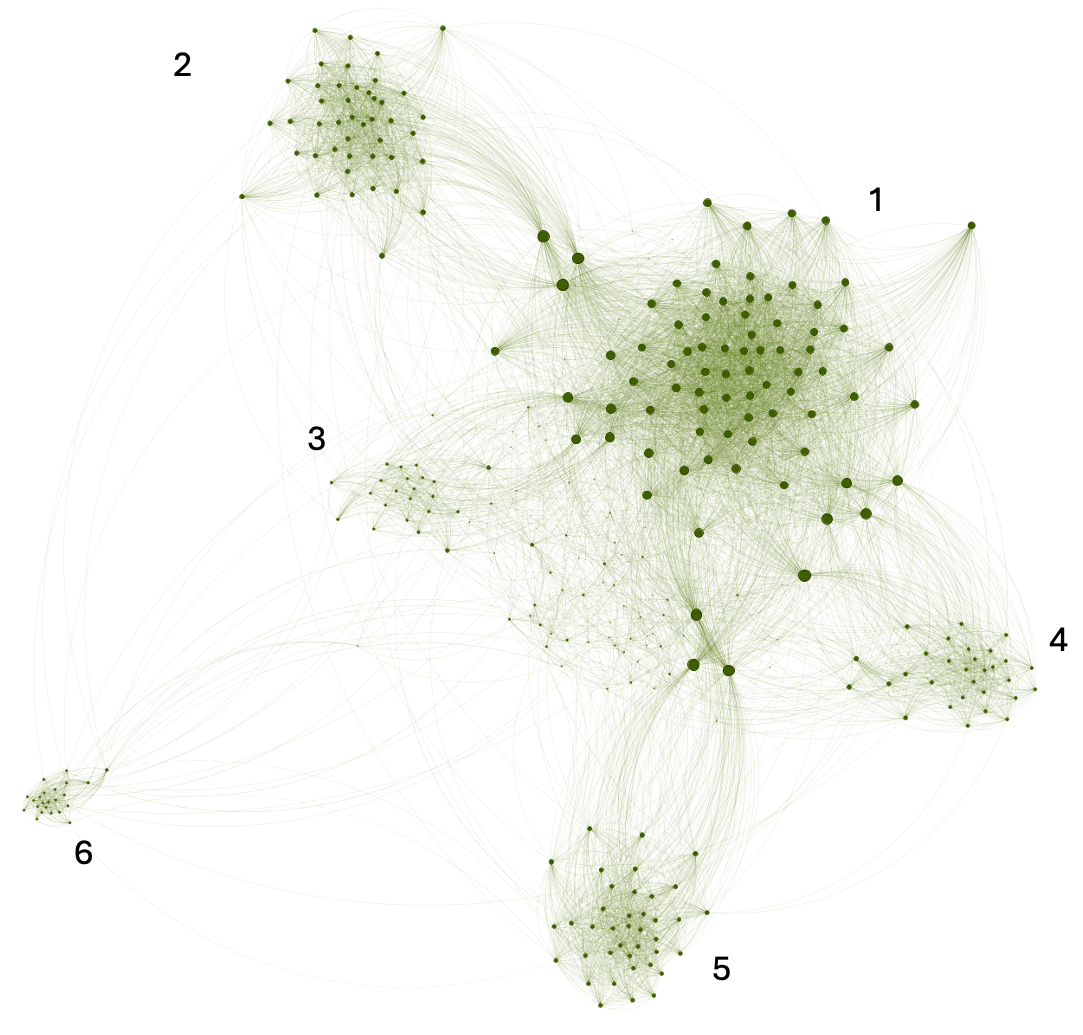}
    \caption{Network of coordinated accounts based on co-retweet activity.}
    \label{fig:cort_top1pct}
\end{figure}
% Following a method proposed by~\citet{luceri_unmasking_2024}, we construct a network based on shared retweets.
% We split the data into 30-minute windows and draw edges between users if both users retweeted the same tweet in the same time window, after dropping users with fewer than 5 and tweets with fewer than 10 retweets.
% We then apply TF-IDF vectorization on the adjacency matrix and compute cosine similarities between user vectors, which serve as edge weights in the network.
% Based on this approach, nodes with the highest eigenvector centralities show the highest likelihood of coordination.
Following~\citet{luceri_unmasking_2024}, we construct a network based on retweet similarity, or simply co-retweet.
First, we drop users who retweeted $< 5$ times and tweets with $< 10$ retweets, and we split the data into 30-minute time windows.
% Each tweet is tagged with a unique identifier tied to its time window.
For every user, we count how many times they retweet each tweet within that window to form a windowed retweet vector.
We apply TF–IDF to these vectors and then compute cosine similarity between every pair of users.
The result is a user-user network in which edges are weighted by those similarity scores.
% Each tweet is assigned a unique tweet-window ID, such that tweet \textit{i} has identifiers $\{i_{t1}, i_{t2}...{i_{tN}}\}$.
% Then, for each user, we obtain the number of times they have retweeted each tweet in a given time window, such that each user is represented with a windowed retweet vector consisting of concatenated tweet-window IDs.
% We TF-IDF transform these user vectors and then obtain pairwise cosine similarities as edge weights between users.
% Based on this method, nodes with the highest eigenvector centralities have the greatest likelihood of coordination. 
% First, we split the data into 30-minute windows and connect users who retweeted the same tweet within the same window, after removing users with $< 5$ retweets and tweets with $< 10$ retweets.
% We apply TF-IDF vectorization to the adjacency matrix and compute cosine similarities between user vectors as edge weights. 
% Based on this method, nodes with the highest eigenvector centralities have the greatest likelihood of coordination. 
% , although this coordination is not necessarily always inauthentic.

\begin{table*}[t!]
    \small
    \centering
    \begin{tabular}{lp{15cm}}
    \toprule
       \textbf{Subcluster}  & \textbf{Most retweeted post} \\
       \midrule
        1 & New Japanese research pre-print publication \newline \newline `SARS-CoV-2 vaccination was associated with higher risk of myocarditis death, not only in young adults but also in ALL age groups including the elderly' \newline \newline \textit{[links to paper]} \\
        \midrule
        2-5 & You REALLY couldn’t make it up. \newline \newline I repeat \newline \newline `We are dealing with the greatest miscarriage of medical science, attack on democracy, damage to population health and erosion of trust in medicine that we will witness in our lifetime' \newline \newline \textit{[quote-tweets post discussing paper that shows ``negative vaccine effectiveness'']}  \\
        \midrule
        6 & ``Breaking news'' on T-cells \newline \newline \textit{[links to 6 papers, dated between May 2020 and November 2021, that show T-cell immunity after infection with COVID-19]} \\
        \bottomrule
    \end{tabular}
    \caption{Most-retweeted posts per coordinated subcluster.}
    \label{tab:cluster_rts}
\end{table*}

We retain the top 1\% of eigenvector centrality values as coordinated nodes, as per~\citet{luceri_unmasking_2024}.
This threshold is chosen as a conservative cutoff, following~\citet{luceri_unmasking_2024}'s intuition that precision is more important than recall in the context of coordination: it is important not to attribute coordinated activity to non-coordinated actors.
Nonetheless, we verify that the coordinated network structure remains the same with cutoffs of 0.5\% and 2\% (see the Appendix).
The resulting network with a 1\% centrality threshold is shown in Figure~\ref{fig:cort_top1pct}.
There are six distinct subclusters emerging.
To provide more context, we show the most-retweeted post from each cluster in Table~\ref{tab:cluster_rts}.
These posts also align with the highest TF–IDF weights in their respective clusters, indicating the strongest coordination signals.
We also use topic labels based on BERTopic modeling of paper abstracts to quantify the types of papers cited in the most shared tweets per cluster.
Invariably, clusters 1-5 predominantly share papers on vaccines or boosters, viral mutations and immunity, and excess deaths and mortality, in this order. 
Cluster 6 predominantly shares papers on vaccines or boosters, T-cells and cell memory, and excess deaths and mortality.
% Five of the six clusters predominantly focus on vaccine risks or supposed ineffectiveness (with four of these having the same most-retweeted post), while the sixth focuses more on pro-natural immunity.

We perform deeper analyses of the users that the coordinated network promotes, but we defer discussion of this to the Superspreader section below.
Moreover, we characterize them in terms of their profile features and topical focus.

\begin{figure*}[t!]
    \centering
    \begin{subfigure}[t]{0.245\linewidth}
        \centering
        \includegraphics[width=0.99\linewidth]{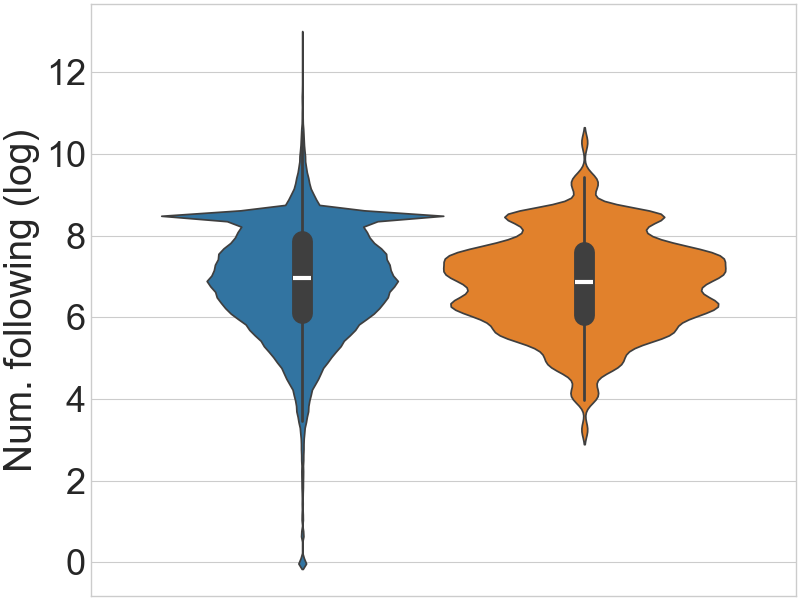}
        \caption{Following}
        \label{subfig:displot_following}
    \end{subfigure}
    \begin{subfigure}[t]{0.245\linewidth}
        \centering
        \includegraphics[width=0.99\linewidth]{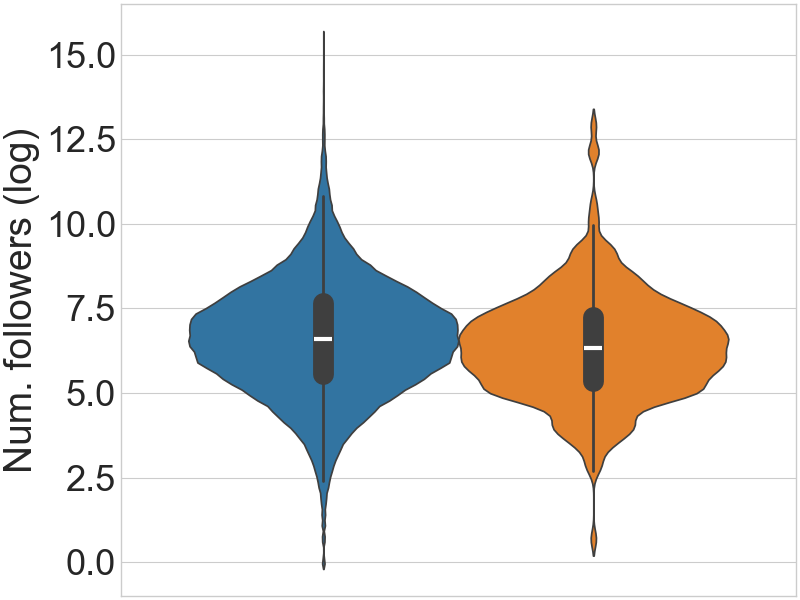}
        \caption{Followers}
        \label{subfig:displot_followers}
    \end{subfigure}
    \begin{subfigure}[t]{0.245\linewidth}
        \centering
        \includegraphics[width=0.99\linewidth]{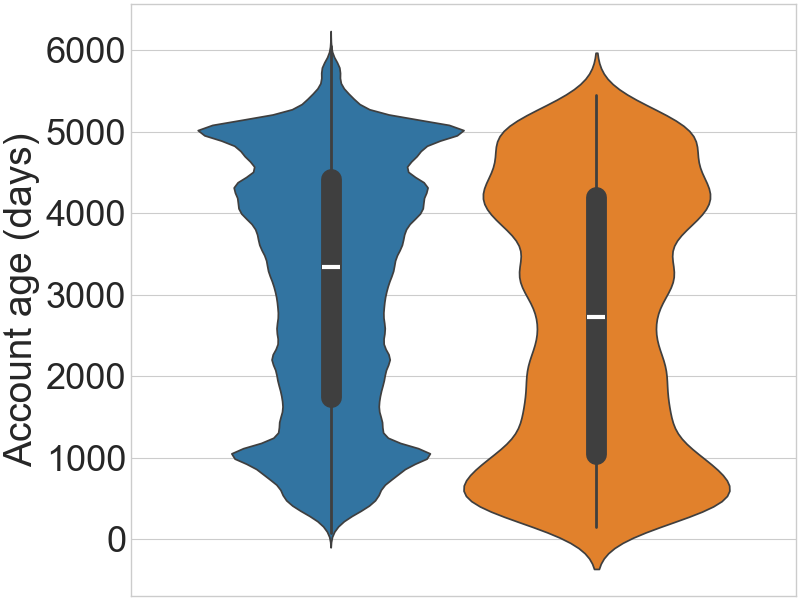}
        \caption{Account age (in days)}
        \label{subfig:displot_age}
    \end{subfigure}
    \begin{subfigure}[t]{0.245\linewidth}
        \centering
        \includegraphics[width=0.99\linewidth]{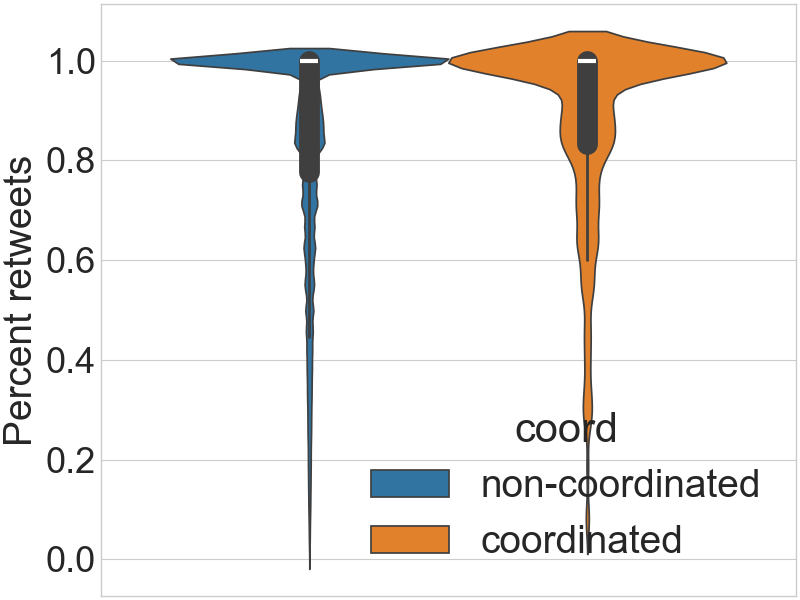}
        \caption{\% Retweets}
        \label{subfig:displot_pct_rts}
    \end{subfigure}
    \caption{Violin plots for distribution comparisons between coordinated and non-coordinated accounts.}
    \label{fig:displots}
\end{figure*}

\paragraph{Profiles.}
To derive coordinated account characteristics, we perform bootstrapped independent samples t-tests across 10,000 iterations to compare the coordinated accounts with random samples of equal size (\textit{N} = 306 per group).
This random sampling is done only from the set of non-coordinated users (i.e., users \textit{not} in the top 1\% of centrality values) with $\geq 5$ retweets to maintain consistency with the coordinated set.
We opt for bootstrapping rather than direct comparisons to the non-coordinated population, as the size of the population can inflate statistical significance; the statistics we report are averages across the bootstrapped iterations.
Coordinated accounts, on average, follow and are followed by a similar number of accounts, and have similar percentages of their science-related tweets being retweets (all bootstrapped $p > 0.1$) as the non-coordinated samples.
However, coordinated accounts are significantly younger in terms of days since they were created ($t = -3.25, p = 0.001$).
We show the distributions of these metrics in Figure~\ref{fig:displots}.
Distributions for the non-coordinated accounts are based on the entire non-coordinated population.

Nearly all (96.4\%) of the coordinated accounts are contrarian, compared to only about 29\% of accounts in the non-coordinated population.
This discrepancy is significant in a chi-squared test ($\chi^2_{(1)}=672.37, p < 0.001$).
% for the contingency table).
For this test, we estimate the expected number of contrarians and conformists in the coordinated network by obtaining their proportions from the non-coordinated population and scaling them down to the coordinated network’s size (rounded to the nearest integer). We then compare these expected values with the observed counts in the coordinated network (Table~\ref{tab:chi_contingency}).
% For this test, we draw the percentages of contrarians and conformists from the non-coordinated population and derive the \textit{expected} number of contrarians and conformists (to the nearest integer) given the coordinated network's sample size.
% This is then compared to the \textit{observed} number of contrarians and conformists in the coordinated network (Table~\ref{tab:chi_contingency}).

\begin{table}[t!]
    \centering
    \begin{tabular}{lrr}
    \toprule
        \textbf{Group} & \textbf{Expected} & \textbf{Observed} \\
        \midrule
        \textbf{Contrarian} & 89 & 295 \\
        \textbf{Conformist} & 217 & 11 \\
        \bottomrule
    \end{tabular}
    \caption{Contingency table for coordination vs stance.}
    \label{tab:chi_contingency}
\end{table}

\paragraph{Activity.}

Next, we focus on characterizing coordinated accounts' retweeting activity, as this is the signal we used to detect coordination; 59.2\% of coordinated accounts only ever perform retweet actions on COVID-19 scientific discussions, compared to 30.6\% in the overall population with five or more retweets ($\chi^2_{(1)} = 116.22, p < 0.001$, medium effect $\phi = 0.44$).
Of the retweeted accounts, 75.9\% of them are contrarian, while 90.4\% of the tweets retweeted are from these contrarian accounts.
Overall, retweets by the coordinated network most often discuss papers on vaccines or boosters (33.8\%), virus mutations and natural immunity (18.4\%), or excess deaths and mortality (9.5\%).
These top-retweeted topics are similar to the topics that are most prominent in the overall discussion; however, the distribution is more unequal and concentrated among the top few in the coordinated network (Gini coefficient, \textit{G} = 0.80) as opposed to the distribution of the overall network (\textit{G} = 0.69) or the rest of the contrarian subset (\textit{G} = 0.76).
To test for statistical significance in this inequality, we perform 10,000 bootstrapped iterations where we obtain random samples of \textit{N} equal to the size of the coordinated network from the overall and contrarian subsets with $\geq 5$ retweets.
In each iteration, we compute the \textit{G} difference between the coordinated network and the comparison sample, and obtain the mean difference as well as 95\% confidence intervals.
In both the overall ($M_{diff} = 0.11, SE=0.0001$, 95\% CI: [0.1094, 0.1099]) and contrarian subset ($M_{diff} = 0.054, SE=0.0001$, 95\% CI: [0.0534, 0.0538]) comparisons, the coordinated network demonstrates significantly higher inequality (i.e., the 95\% CIs do not span zero).
That is, coordinated accounts focus on a narrower set of topics than the rest of the network.
% , as the Gini coefficient for topic share percentage is higher among coordinated accounts (Figure~\ref{fig:topic_ecdf}).

% \begin{figure}
%     \centering
%     \includegraphics[width=0.99\linewidth]{figures/topic_share_ecdf.png}
%     \caption{Empirical Cumulative Distribution Function (ECDF) of topic share percentages with Gini coefficients.}
%     \label{fig:topic_ecdf}
% \end{figure}

\subsection{Superspreaders}

To detect superspreaders, we follow the approach outlined by~\citet{deverna_identifying_2024}, who propose an h-index metric for uncovering influential spreaders.
Accordingly, superspreaders are defined as users with a high number of posts that also tend to get a high number of retweets; for example, a user with a superspreader score (i.e., h-index) of 10 is a user with at least 10 tweets that received at least 10 retweets \textit{each}.
We consider accounts in the top 1\% of scores as superspreaders, in line with~\citet{deverna_identifying_2024}.

\begin{figure*}[t!]
    \centering
    \begin{subfigure}[t]{0.245\linewidth}
        \centering
        \includegraphics[width=0.99\linewidth]{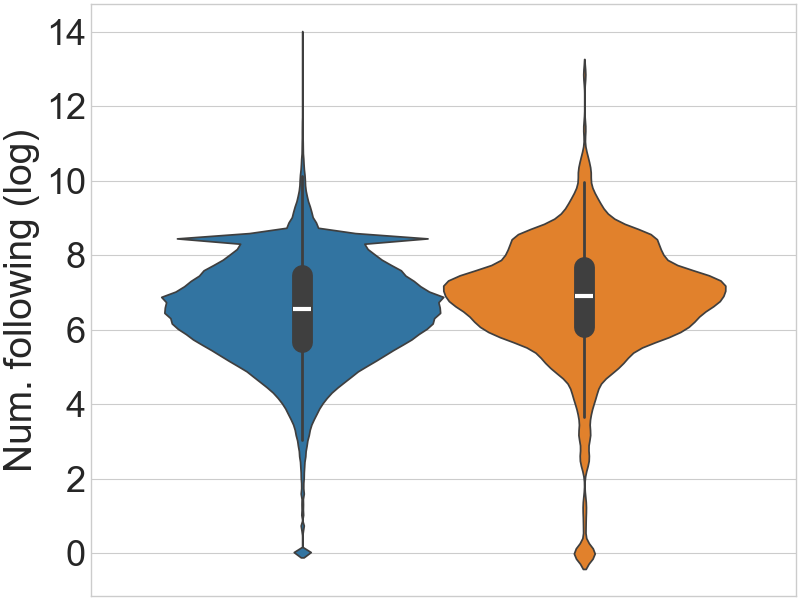}
        \caption{Following}
        \label{subfig:dispsp_following}
    \end{subfigure}
    \begin{subfigure}[t]{0.245\linewidth}
        \centering
        \includegraphics[width=0.99\linewidth]{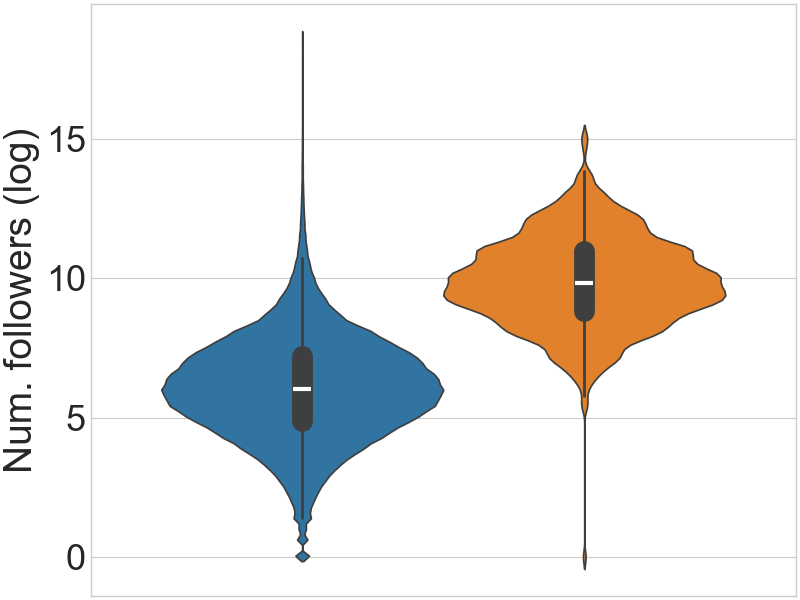}
        \caption{Followers}
        \label{subfig:dispsp_followers}
    \end{subfigure}
    \begin{subfigure}[t]{0.245\linewidth}
        \centering
        \includegraphics[width=0.99\linewidth]{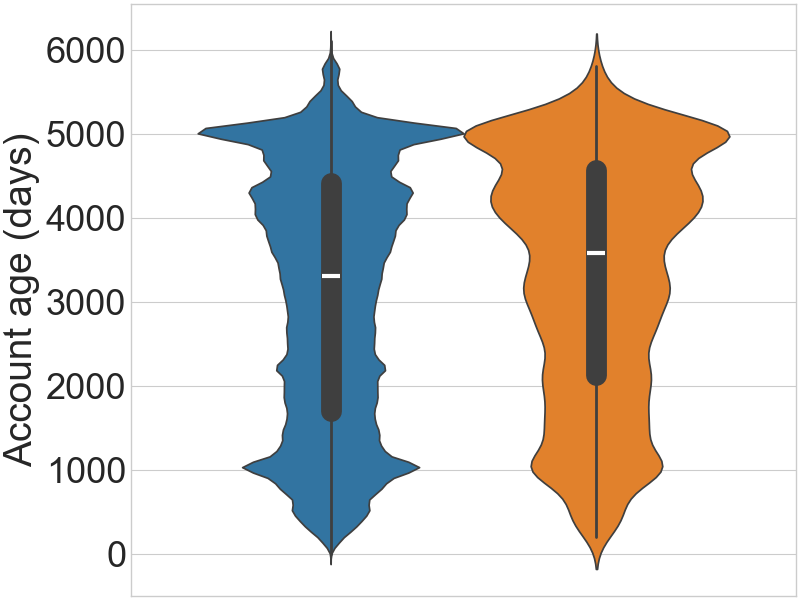}
        \caption{Account age}
        \label{subfig:dispsp_age}
    \end{subfigure}
    \begin{subfigure}[t]{0.245\linewidth}
        \centering
        \includegraphics[width=0.99\linewidth]{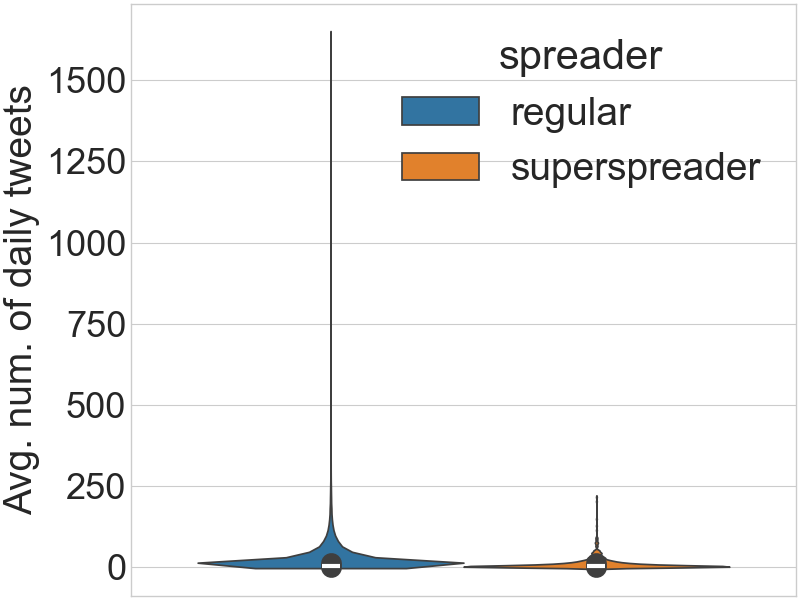}
        \caption{Daily tweets}
        \label{subfig:dispsp_avg_daily}
    \end{subfigure}
    \caption{Violin plots for distribution comparisons between superspreader and non-superspreader characteristics.}
    \label{fig:disp_spreaders}
\end{figure*}

\paragraph{Profiles.}
As with the coordinated network, we first characterize the superspreader accounts by comparing them against bootstrapped random samples of equal size (\textit{N} = 764) using independent-samples t-tests.
We plot the distributions of the account characteristics we compare in Figure~\ref{fig:disp_spreaders}.
Superspreaders tend to follow more accounts overall ($t = 3.96, p < 0.001$), this being driven by the tail-ends of the distribution (Figure~\ref{subfig:dispsp_following}).
They are also slightly older accounts ($t = 3.05, p = 0.013$) but do not differ in terms of daily volume of tweets ($t = -1.90, p = 0.12$).
Moreover, we observe that superspreaders have significantly more followers than the general sample ($t = 42.71, t < 0.001$), with a clear mean difference in their distribution (Figure~\ref{subfig:dispsp_followers}).

Superspreaders also tend to be (legacy) verified at higher rates (28\% vs. only 1.8\% in the general population), and they are more likely to be conformists (80.6\% vs. 70.8\% in the general population; $\chi^2 = 35.62, p < 0.001$).
Their substantially higher number of followers, higher verification rates, and more pro-science stance suggest that superspreaders of COVID-19 science are predominantly science communicators or science professionals. % as opposed to malicious actors.
Our manual assessment of these accounts is in line with this observation (see Appendix for our labeling approach and criteria).
As seen in Table~\ref{tab:xtab_credentials}, the majority of conformist superspreaders are physicians/medical doctors or scientists actively working in COVID-related fields, with a substantial proportion of contrarian superspreaders (39.8\% jointly) having similar credentials.
The largest class of contrarian superspreaders, however, is people who do not primarily work within the medical science field, though they are often experts in other fields (e.g., law, economics, etc. -- see the Appendix for how the scientific credentials in Table~\ref{tab:xtab_credentials} are defined).

\begin{table*}[t]
    \centering
    \begin{tabular}{lrrrrrr}
    \toprule
        \textbf{Type} & \textbf{Physician} & \textbf{Scientist} & \textbf{Science communicator} & \textbf{Science organization} & \textbf{Other/non-credentialed} \\
        \midrule
        \textbf{Conformist} & 166 (26.9\%) & \textbf{207 (33.6\%)} & 41 (6.65\%) & 83 (13.47\%) & 119 (19.3\%) \\
        \textbf{Contrarian} & 31 (20.9\%) & 28 (18.9\%) & 3 (2.02\%) & 4 (2.70\%) & \textbf{82 (55.4\%)} \\
        \bottomrule
    \end{tabular}
    \caption{Cross-tabulation of scientific credentials by superspreader type.}
    \label{tab:xtab_credentials}
\end{table*}

\paragraph{Topics.}
% We analyze the topics of papers discussed by conformist and contrarian superspreaders separately, as these groups may engage distinct audiences, i.e., conformist superspreaders may be engaging in more general science communication, while contrarian superspreaders may be focusing on specific topics that more effectively indict official institutions.
% Indeed, conformist superspreaders have a more equal topic share distribution (\textit{G} = 0.58) than contrarian superspreaders (\textit{G} = 0.74).
% Moreover, conformists' topics are more equal, and contrarians' more unequal, than the overall network (\textit{G} = 0.68).
% Therefore, conformist superspreaders' topical activity is more spread out than contrarians, who tend to focus on fewer topics.
We analyze the topics of papers discussed by conformist and contrarian superspreaders separately, as these groups may engage distinct audiences: conformist superspreaders may participate in broader science communication, whereas contrarian superspreaders may concentrate on topics that more directly challenge or indict official institutions. Consistent with this interpretation, conformist superspreaders exhibit a more even distribution of topic engagement (\textit{G} = 0.58) than contrarian superspreaders (\textit{G} = 0.74). Moreover, conformists’ topical activity is more evenly distributed---and contrarians’ more concentrated---than that of the overall network (\textit{G} = 0.68). Together, these results indicate that conformist superspreaders engage across a wider range of scientific topics, while contrarian superspreaders focus on a narrower subset.
% Indeed, as seen from the distributions and accompanying Gini coefficients in Figure~\ref{fig:sp_topic_share}, conformist superspreaders have a more equal topic distribution (lower Gini), while contrarians have a more unequal one (higher Gini) than the overall paper topic distribution.
% This offers some evidence for our earlier intuition.

% \begin{figure}
%     \centering
%     \includegraphics[width=0.99\linewidth]{figures/superspreader_topics_ecdf.png}
%     \caption{ECDF of topic share distributions by superspreader status.}
%     \label{fig:sp_topic_share}
% \end{figure}

In terms of the most discussed topics, vaccines/boosters and epidemic models are the top two for both conformists (9.4\% and 9.1\%, respectively) and contrarians (14.1\% and 12.7\%, respectively).
For conformists, the top five are completed with genomic sequencing/mutations (7.7\%), viral mutations/natural immunity (7.3\%), and seroprevalence (i.e., how widespread the virus is; 7.2\%).
For contrarians, the top five are completed by excess mortality (11.6\%), vaccination coverage/immunity (8.4\%), and seroprevalence (8.3\%).
The distribution of topical prevalence for the top 20 topics per group is shown in the Appendix.
% We show the distribution of topical prevalence per group and overall for their respective top 20 topics in the Appendix.

\paragraph{Key terms.}

\begin{figure}[t!]
    \centering
    \begin{subfigure}[t]{0.49\linewidth}
        \centering
        \includegraphics[width=0.99\linewidth]{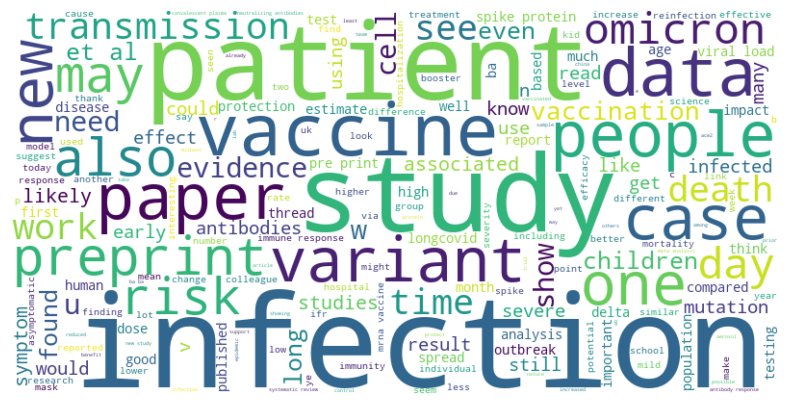}
        \caption{Conformist keywords}
        \label{subfig:conf_words}
    \end{subfigure}
    \begin{subfigure}[t]{0.49\linewidth}
        \centering
        \includegraphics[width=0.99\linewidth]{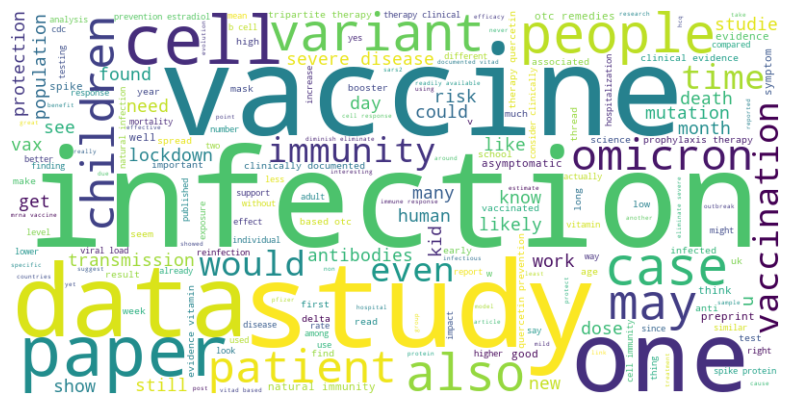}
        \caption{Contrarian keywords}
        \label{subfig:cont_words}
    \end{subfigure}
    \begin{subfigure}[t]{0.49\linewidth}
        \centering
        \includegraphics[width=0.99\linewidth]{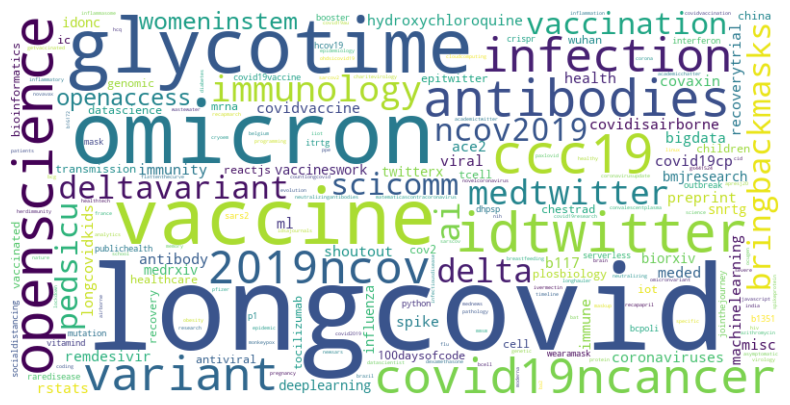}
        \caption{Conformist hashtags}
        \label{subfig:conf_hash}
    \end{subfigure}
    \begin{subfigure}[t]{0.49\linewidth}
        \centering
        \includegraphics[width=0.99\linewidth]{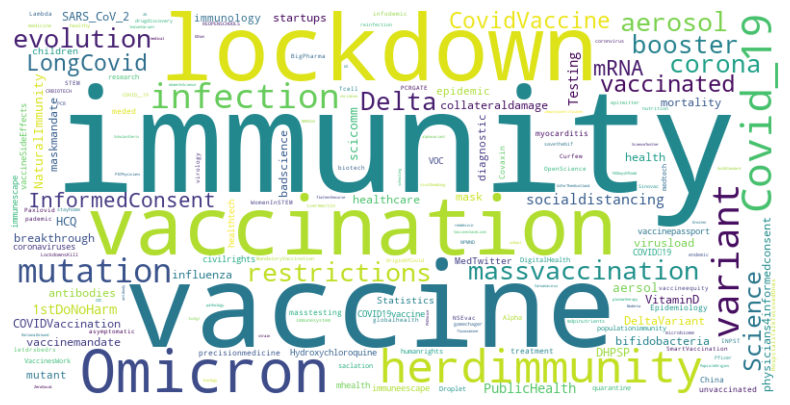}
        \caption{Contrarian hashtags}
        \label{subfig:cont_hash}
    \end{subfigure}
    \caption{Word clouds for main keywords and hashtags by superspreader type.}
    \label{fig:wordclouds}
\end{figure}

To better understand how these topics are discussed across contrarian and conformist superspreaders, we extract the top keywords and top hashtags from the original (i.e., non-retweet) tweets of these groups.
We show these as wordclouds in Figure~\ref{fig:wordclouds}; word size is proportional to the relative frequency across the respective group's tweets.
We exclude words like ``COVID-19'' and other synonyms (e.g., ``coronavirus'', ''SARS-COV-2'' etc.) due to their topical prevalence.

The vocabulary between contrarian and conformist superspreaders is fairly similar, although we see that contrarians pay greater focus to a select few words (vaccine, infection, study), while conformists use a broader range of keywords.
Ranking these words in terms of their frequency per group, we find a moderately strong rank correlation between the two groups (Kendall $\tau = 0.57, p < 0.001$).

However, in terms of hashtags, there are more stark contrasts.
Whereas conformists are more focused on long COVID in addition to discussion of antibodies, infection, and open science, contrarians use hashtags associated with more contentious COVID-19 topics like (herd) immunity, vaccination, and lockdowns.
This is also reflected in the relatively lower rank correlation in hashtag frequency between conformist and contrarian superspreaders (Kendall $\tau = 0.38, p < 0.001$), which is nevertheless also statistically significant.

\paragraph{Emotions.}

An important aspect in characterizing the detected superspreaders is the emotionality expressed in their posts, as this can offer crucial context on the potential intent in sharing science on Twitter.
We detect emotions of (non-retweet) posts by superspreaders and compare them against an equally large, randomly drawn sample of posts.
We use the DistilRoBERTa model by~\citet{hartmann2022emotionenglish}, and take the majority label for each post.
We show results in Table~\ref{tab:super_emotions}.

\begin{table*}[t!]
    \centering
    \begin{tabular}{lrrrrrrr}
    \toprule
        \textbf{User type} & \textbf{Anger} & \textbf{Disgust} & \textbf{Fear} & \textbf{Joy} & \textbf{Neutral} & \textbf{Sadness} & \textbf{Surprise} \\
        \midrule
        Random contrarian & \textbf{7.66} & 0.99 & 37.93 & 5.60 & 32.30 & \textbf{6.08} & \textbf{9.44} \\
        Random conformist & 3.62 & 0.61 & 37.50 & \textbf{7.29} & 40.47 & 4.20 & 6.31 \\
        Superspreader contrarian & 4.22 & \textbf{0.99} & 35.31 & 3.34 & \textbf{44.34} & 4.40 & 7.39 \\
        Superspreader conformist & 2.57 & 0.44 & \textbf{50.41} & 3.78 & 35.31 & 3.16 & 4.32 \\
        \bottomrule
    \end{tabular}
    \caption{Percentage of emotion share per user type.}
    \label{tab:super_emotions}
\end{table*}

Overall, this analysis shows that the most valenced group (least neutral) is the randomly sampled contrarians, as opposed to the contrarian superspreaders, who are the most neutral.
The former group makes the most extensive use of negative emotions, such as anger and sadness.
Conformist superspreaders demonstrate by far the most fear in their posts, possibly as an artifact of the warning-style posting they employ.

\begin{figure}[t!]
    \centering
    \begin{subfigure}[t]{0.99\linewidth}
        \centering
        \includegraphics[width=0.99\textwidth]{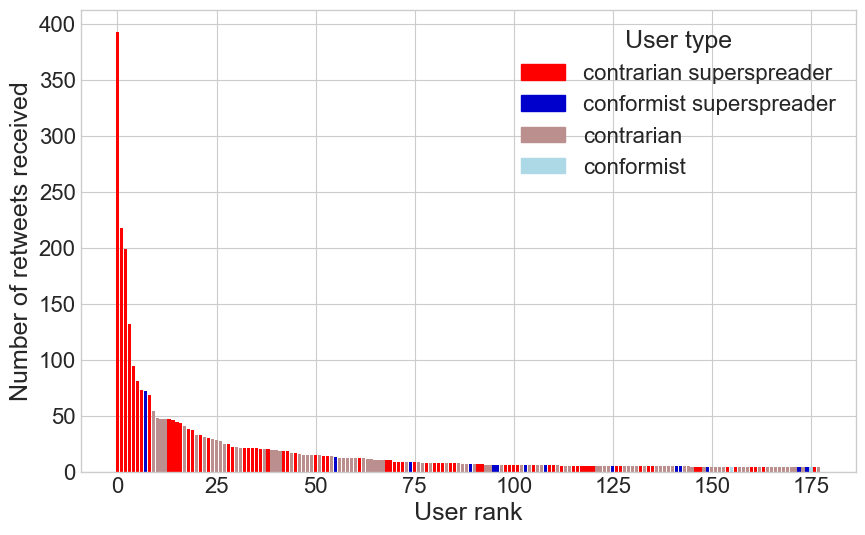}
        \caption{Coordinated network.}
        \label{subfig:coord_sp_bar}
    \end{subfigure}
    \begin{subfigure}[t]{0.99\linewidth}
        \centering
        \includegraphics[width=0.99\linewidth]{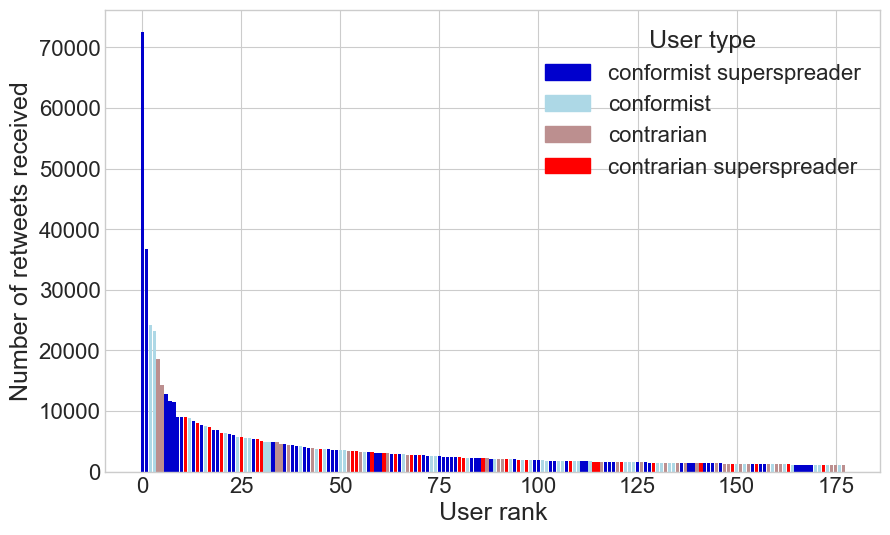}
        \caption{Rest of the network.}
        \label{subfig:ovrl_sp_bar}
    \end{subfigure}
    \caption{Distribution of most-retweeted accounts by superspreader and related stance.}
    \label{fig:sp_distros}
\end{figure}

\paragraph{Coordinated amplification of superspreaders.}
To determine whether there is cross-activity between different types of platform actors, we also perform an analysis of whether accounts in the coordinated retweet network tend to promote these detected superspreaders.
Overall, among the users who are retweeted by the coordinated network, we find 14.6\% of the conformist superspreaders and 78.4\% of the contrarian superspreaders.
The majority of retweets amplify contrarian superspreader accounts (51.0\%) followed by other contrarian accounts (39.4\%).
Only a small minority of retweets are for conformist superspreaders (6.21\%) or other conformists (3.40\%).
This is demonstrated in the bar plot in Figure~\ref{subfig:coord_sp_bar}, where we show the distribution of retweets per user who received more than 3 retweets in the coordinated network.
By comparison, the most-retweeted accounts in the rest of the network tend to be conformist superspreaders (Figure~\ref{subfig:ovrl_sp_bar}).

Importantly, the accounts promoted by the coordinated network are not the most popular overall, even when only considering contrarian superspreaders (see Table~\ref{tab:coord_sup_descs}, which shows the relative ranking of the top-5 most promoted superspreaders).
%, along with a brief description of their profiles.
Moreover, the rank correlation of the most-retweeted accounts between the coordinated and general network (Kendall $\tau = 0.14, p < 0.001$) is low.
Looking at the top 884 most-retweeted accounts overall (884 is chosen as this is the number of accounts retweeted within the coordinated network), 576 of them are never retweeted by the coordinated network.
This pattern also holds when only considering contrarian accounts.
The rank correlation remains low-to-moderate (Kendall $\tau = 0.38, p < 0.001$), but 449 of the 884 most-retweeted contrarians overall are not retweeted by the coordinated network.
Thus, this shows that the coordinated network focuses on a specific subset of accounts that are not necessarily the most popular in the social network.

\begin{table*}[t!]
    \small
    \centering
    \begin{tabular}{rrr|l}
    \toprule
        \multicolumn{3}{c}{\textbf{Rank}} & \\
        \midrule
        \textbf{Coordinated network} & \textbf{Overall network} & \textbf{Contrarian subset} & \textbf{Description} \\
        % $R_p$ & $R_{ovrl}$ & $R_c$ & \textbf{Description} \\
        \midrule
        1 & 385 & 82 & Anti-vaccine cardiologist \\
        2 & 149 & 30 & Anti-lockdown professor \\
        3 & 37 & 9 & Pro-herd immunity pathologist \\
        4 & 149 & 30 & Anti-vaccine physician \\
        5 & 122 & 26 & Pro-herd immunity physician \\
        \bottomrule
    \end{tabular}
    % \caption{Relative ranks of most-promoted superspreaders. Ranks refer to, in order of columns, the coordinated network-promoted rank as in Figure~\ref{fig:coord_sp_bar}, superspreader scores overall, and superspreader scores among contrarians only.}
    \caption{Ranks refer to the most-retweeted accounts in each set, with 1 being the highest.}
    \label{tab:coord_sup_descs}
\end{table*}

\subsection{Bots}

For accounts with ten COVID-19 related tweets and above (\textit{N} = 28.9k), we also collect their bot scores using Botometer's historical data; this is suitable for our dataset as it was collected before Botometer stopped operating due to Twitter API changes on May 31st, 2023.\footnote{\url{https://botometer.osome.iu.edu/}}

\begin{figure}[t!]
    \centering
    \begin{subfigure}[t]{0.49\linewidth}
        \centering
        \includegraphics[width=0.99\linewidth]{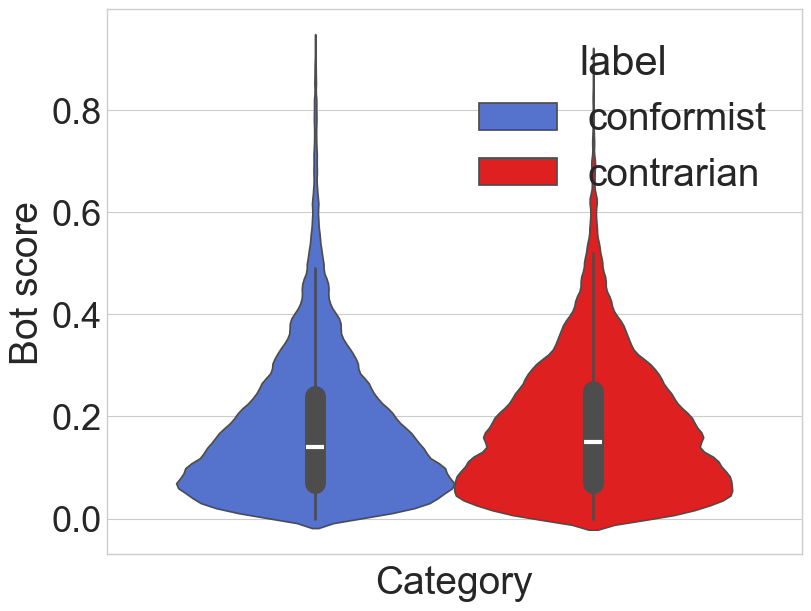}
        \caption{By account stance}
        \label{subfig:bot_label}
    \end{subfigure}
    \begin{subfigure}[t]{0.49\linewidth}
        \centering
        \includegraphics[width=0.99\linewidth]{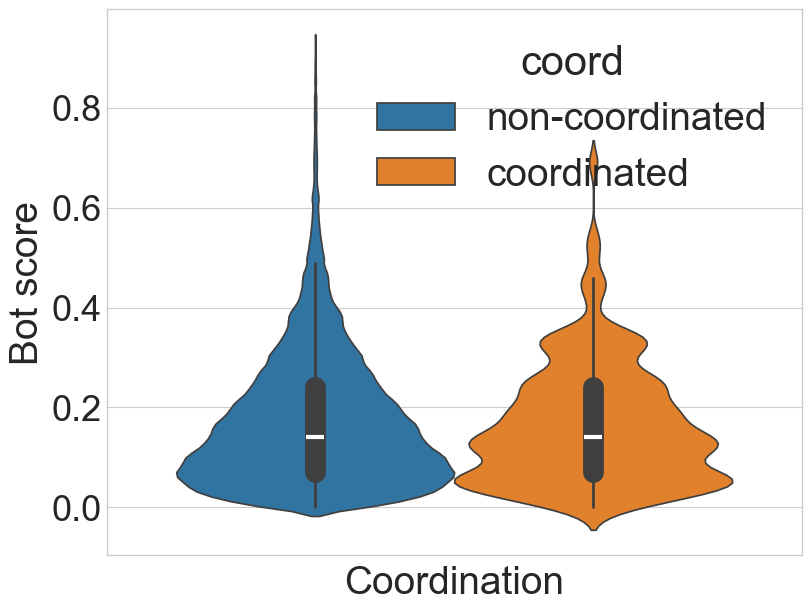}
        \caption{By coordination status}
        \label{subfig:bot_coord}
    \end{subfigure}
    \caption{Bot score distributions.}
    \label{fig:disp_bots}
\end{figure}

This analysis does not yield any suspicious patterns.
Bot scores do not produce significant differences between conformist and contrarian accounts ($t = -1.55, p = 0.12$; Figure~\ref{subfig:bot_label}) nor between coordinated accounts and bootstrapped non-coordinated samples ($t = -0.52, p = 0.53$; Figure~\ref{subfig:bot_coord}).
Indeed, based on a manual inspection, many of the bots appear to be overt ones that are automated to share papers, science news, etc.
For context, from the accounts in the top 1\% of assigned bot scores, just over 1 in 6 (17.8\%) include one of the following overt terms in their name: bot, science, paper, hub, medrxiv, biorxiv, preprint, or article.
These terms are typical in overt, automated, science-sharing bots.
On the other hand, less than 1\% of the accounts in the bottom 1\% of bot scores contain any of those terms in their usernames.
%Thus, in the context of COVID-19 science sharing, 
This finding is consistent with prior work showing that coordination does not necessarily imply automation.

\subsection{Summary (RQ1)}

Taken together, these patterns suggest mostly contrarian coordination.
That is, while superspreaders are predominantly science communicators who promote COVID-19 science to their wider networks, a cluster of coordinated accounts tends to promote mostly anti-consensus superspreaders who share contrarian science, often focusing on a few specific topics.

Automated activity is mostly restricted to overt bots that typically share scientific papers, preprints, or other resources in a transparent manner.
In short, coordination is not driven by automation in this case.
Indeed, several prior works have made the case that coordination and automation are two distinct behaviors, since coordination can also be the result of trolling efforts, human-operated information campaigns, or organic cooperation~\cite{mazza_investigating_2022,nizzoli_coordinated_2021}.
Similarly, coordination alone does not necessarily constitute evidence of malicious intent~\cite{pacheco_uncovering_2021}, however, it can artificially amplify content and create illusions of public consensus~\cite{luceri_unmasking_2024}.
Thus, while we cannot determine the intent behind this coordination, our results suggest a non-automated amplification of specific contrarian accounts.

\section{Media Information Pathways}\label{sec:media}

In our subsequent analyses, we consider the role of news media in propagating COVID-19 science through news articles published in various outlets.
Beyond the media's isolated role, we are also interested in how media outlets' activity is related to the activity of Twitter, and in particular, superspreaders and ``science influencers.''
We consider news articles by media outlets that published COVID-related articles in the English language, and drop any news aggregators (e.g., Yahoo News, Google News) from our analyses.

\subsection{Media Behavior Overview}

\begin{figure}[t!]
    \centering
    \begin{subfigure}[t]{0.99\linewidth}
        \centering
        \includegraphics[width=0.99\linewidth]{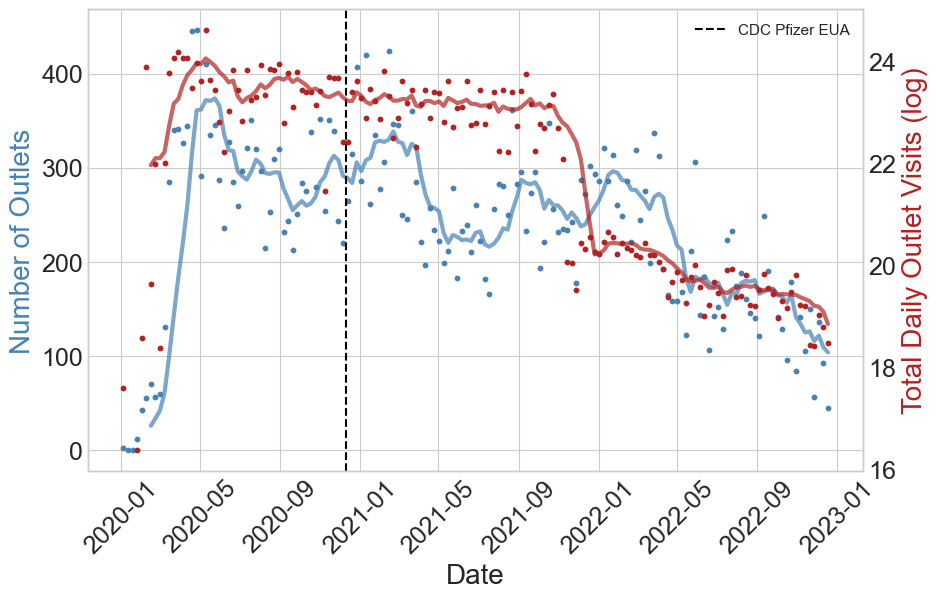}
        \caption{Number and total daily visits of outlet websites.}
        \label{subfig:outlet_visits}
    \end{subfigure}
    \begin{subfigure}[t]{0.99\linewidth}
        \centering
        \includegraphics[width=0.99\linewidth]{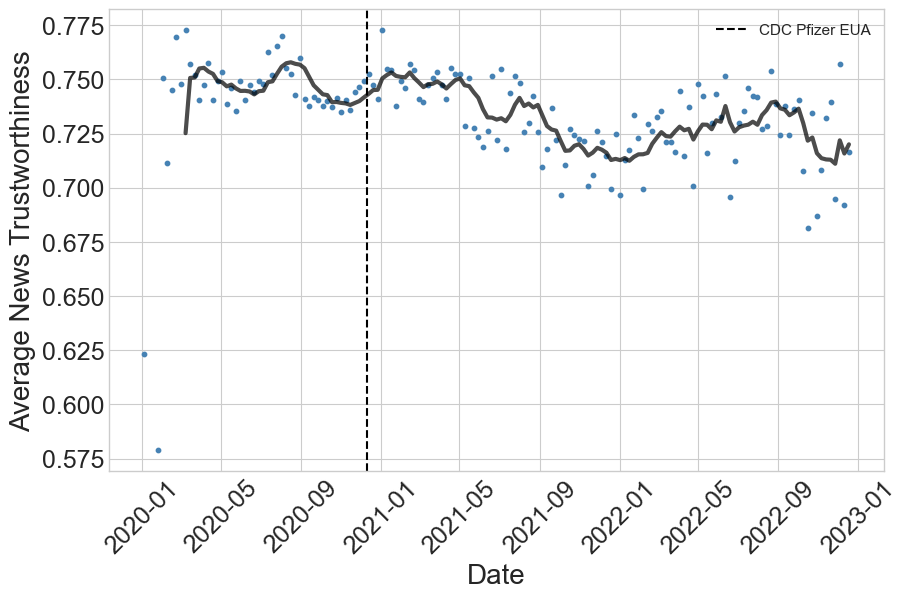}
        \caption{Trust score of outlets.}
        \label{subfig:outlet_trust}
    \end{subfigure}
    \caption{Seven-day rolling averages for number, total daily visits, and average news trustworthiness of news outlets reporting on COVID-19 papers.}
    \label{fig:outlet_activity}
\end{figure}

We start with a broad description of media activity over the observation period.
For these analyses, we rely on the domain trustworthiness scores from~\citet{lin_high_2023}, who combine six expert ratings from media watchdogs, fact-checkers, and researchers, and use imputation to extend coverage to 11.5k news domains.
% For these analyses, we use the domain trustworthiness scores reported by~\citet{lin_high_2023} who utilize six different sets of expert news trustworthiness ratings from media watchdogs, fact checkers, and researchers, and perform imputation to expand scores to 11.5k news domains.
Moreover, we scrape the daily number of visitors for each outlet's website from StatsCrop.
We plot the 7-day rolling averages of the number, popularity, and trust score of media outlets reporting on COVID-19 scientific articles in Figure~\ref{fig:outlet_activity}.
%, which shows an overall downward trend across all metrics.

% Moreover, we observe some syndication patterns.
% Syndication refers to media behavior where different outlets publish the same exact article, either at the same time or very close to one another.
% This is standard in journalism, and may happen in situations where different outlets report the same story after a media embargo is lifted, or cases where large, national outlets license their articles to smaller, local news for more community outreach.
% We report an analysis on syndication in the Appendix, which does not indicate any suspicious media syndication patterns.
% In subsequent analyses, we therefore remove these syndication activities by keeping only the earliest version of a news article.
% Where several articles were published at the \textit{exact} same time, we keep the one by the most popular outlet (by daily visits), operating under the assumption that larger outlets will license their articles to more local ones, rather than vice-versa.

We also observe some syndication patterns. 
Syndication occurs when different outlets publish the same article at the same time or very close to one another. 
This is standard in journalism, often happening when outlets report the same story after a media embargo is lifted, or when large national outlets license articles to smaller local outlets for broader community reach. 
We report an analysis of syndication in the Appendix, which does not indicate any suspicious patterns. 
For subsequent analyses, we remove syndication by keeping only the earliest version of each news article. 
When multiple articles are published at the \textit{exact} same time, we retain the one from the most popular outlet (by daily visits), assuming larger outlets license their articles to smaller ones rather than vice versa. 

\subsection{Crossover of Media and Superspreader Activity}

To assess how media outlets and superspreaders relate to one another, we ask whether they tend to discuss the same papers. We focus on these two actor types because they primarily drive visibility and engagement.
We collect all DOIs shared by either superspreaders or outlets---this set defines our ``vocabulary.''
Each outlet and superspreader is represented as a sparse DOI vector, with values equal to the number of articles (outlets) or tweets (superspreaders) mentioning each paper.
After TF–IDF transformation, we compute cosine similarities between every outlet–superspreader pair, treating similarity as alignment in the papers they discuss.

\begin{figure}[t!]
    \centering
    \begin{subfigure}[t]{0.99\linewidth}
        \centering
        \includegraphics[width=0.99\linewidth]{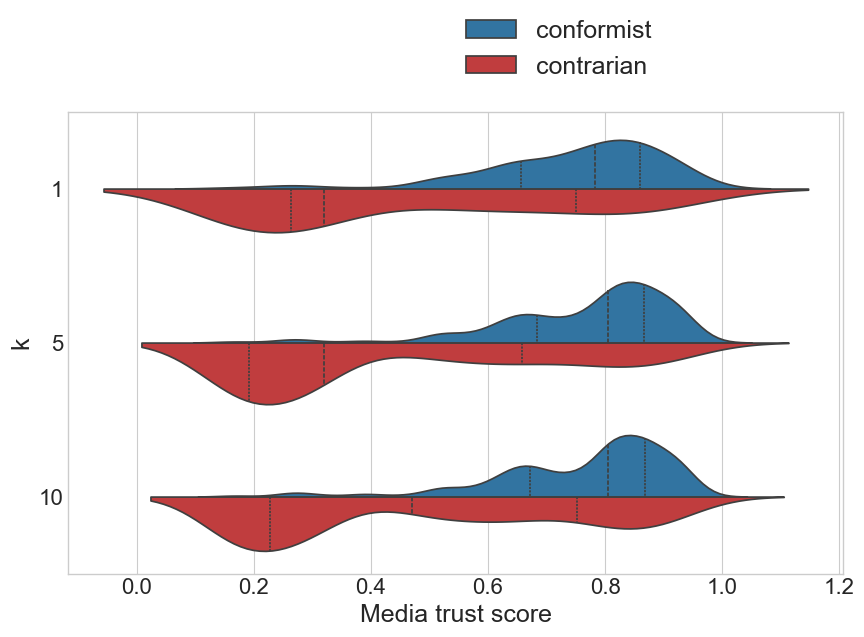}
        \caption{Trust scores}
        \label{subfig:med_trust}
    \end{subfigure}
    \begin{subfigure}[t]{0.99\linewidth}
        \centering
        \includegraphics[width=0.99\linewidth]{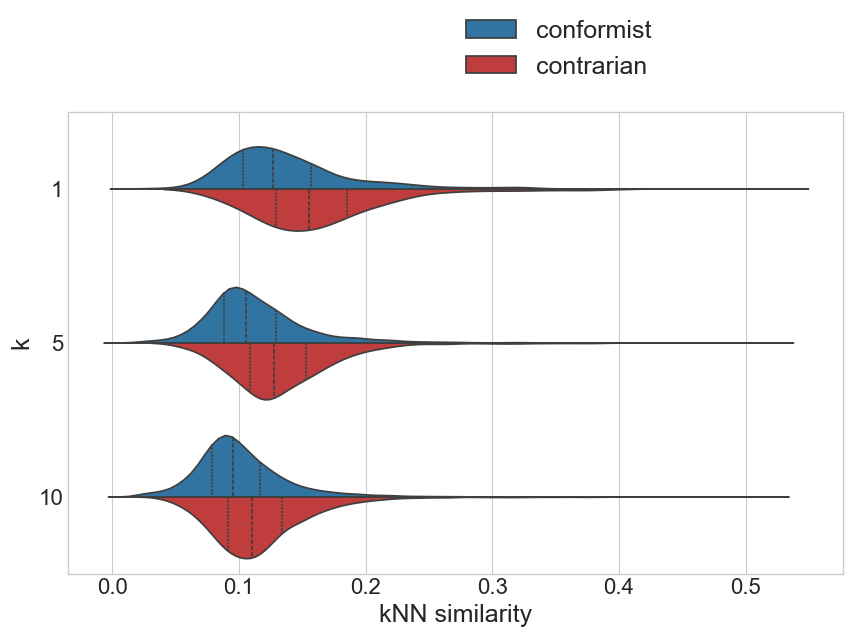}
        \caption{Similarity values}
        \label{subfig:med_sim}
    \end{subfigure}
    \caption{Violin plots representing (a) the trust scores of closest-neighbor media outlets by group, and (b) how similar the activity is between contrarian and conformist superspreaders' closest-neighbor media outlets.}
    \label{fig:media_violin}
\end{figure}

We next identify each superspreader’s K-nearest outlets and compare the neighbors of conformists and contrarians.
We test K = 1, 5, and 10 for robustness, and and show differences between both overall similarities and neighboring outlet trust scores between the two groups in Figure~\ref{fig:media_violin}.
% We then derive each superspreader's K-nearest outlets and compare the neighbors of conformist and contrarian superspreaders.
% For robustness, we evaluate 1, 5, and 10 nearest-neighbors, and show differences between both overall similarities and neighboring outlet trust scores between the two groups in Figure~\ref{fig:media_violin}.

Across all K, contrarian superspreaders align with significantly lower-trust outlets, and are significantly more similar to their outlet neighbors (all $\textit{p} < 0.001$ in independent-samples t-tests).
In short, contrarian superspreaders show stronger alignment with low-credibility outlets than conformists do with high-credibility outlets.

% Across all K-configurations, contrarian superspreaders align with significantly lower-trust outlet neighbors, while having significantly higher similarity values to their neighboring outlets compared to conformists (all $\textit{p} < 0.001$ in independent-samples t-tests).
% In other words, contrarian superspreaders demonstrate stronger alignment with low-credibility outlets, compared to the alignment between conformist superspreaders and high-credibility outlets.

\paragraph{Breakdown of neighboring outlets.}

\begin{figure*}
    \centering
    \includegraphics[width=0.80\linewidth]{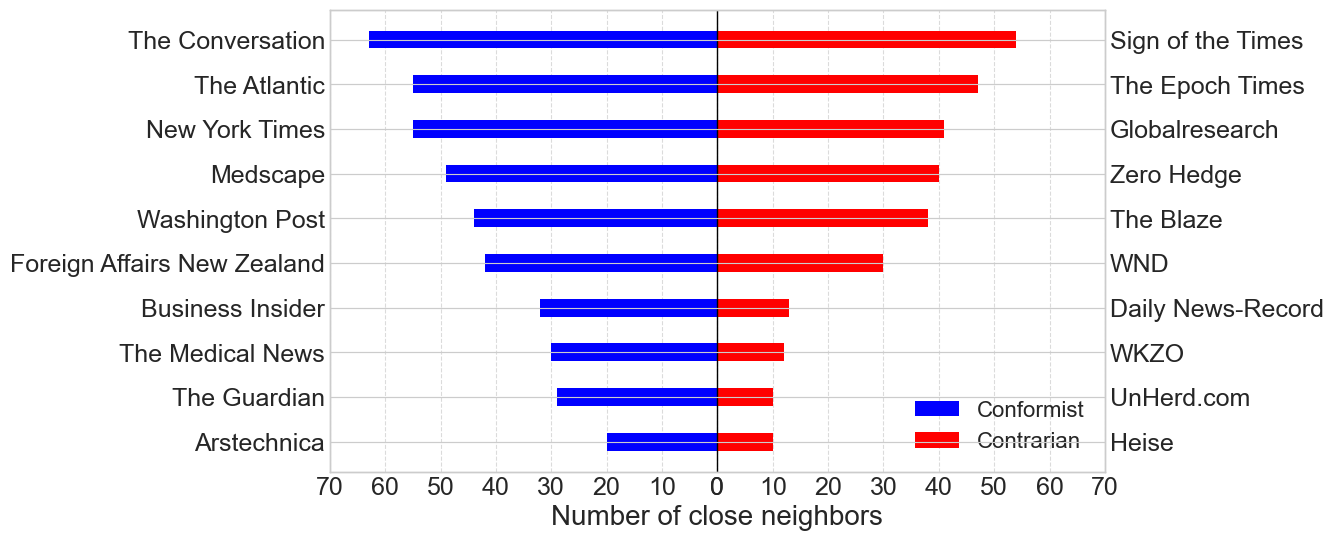}
    \caption{Most frequent closest-neighbor outlets by superspreader group.}
    \label{fig:outlet_breakdown}
\end{figure*}

% Given the similarity in the patterns we observe, we continue the analysis with k=5 as a good middle ground (to avoid over-reliance on a single neighbor, or neighbors whose similarity values are lower).
For further context, we show the news outlets that are most often identified as 5-NN neighbors of conformist and contrarian superspreaders in Figure~\ref{fig:outlet_breakdown}, which confirms the trust patterns we observe in Figure~\ref{fig:media_violin}.
Conformist superspreaders tend to share similar DOIs with mainstream or medical news media, whereas contrarian superspreaders mostly neighbor with conspiratorial, low-quality, or pseudoscientific sources (as determined by Media Bias/Fact Check.\footnote{\url{https://mediabiasfactcheck.com/}})

% To offer a more in-depth view into these relationships, we show the derived network when keeping the TF-IDF weighed edges between superspreaders and outlets in Figure~\ref{fig:med_sp_network}.
% The colored neighborhoods represent the primary conformist and contrarian clusters. 

% \begin{figure}
%     \centering
%     \includegraphics[width=0.85\linewidth]{figures/media_sp.png}
%     \caption{Superspreader-Outlet 5-NN network. Blue shows conformist-mainstream media and red shows contrarian-low trust media neighborhoods.}
%     \label{fig:med_sp_network}
% \end{figure}

\paragraph{Posting times.}

The preceding analyses show how different groups of media outlets and superspreaders talk about different academic papers, but they do not unpack whether one type of source precedes another.
To address this, we examine periods of high activity for each paper DOI.
We first retain only DOIs with at least 100 mentions in both media outlets \textit{and} Twitter. 
% First, we filter such that we only retain paper DOIs with at least 100 mentions in media outlets \textit{and} Twitter.
% out any paper DOIs with fewer than 100 mentions in media outlets \textit{or} Twitter. 
For each DOI, we apply Gaussian kernel density estimation on all mentions by news outlets and superspreaders between the DOI's first and last mention. 
We then scan the density grid to identify the smallest interval that contains at least 50\% of total activity.
% We then iterate through the density grid to identify the smallest region containing \textit{at least} 50\% of the density (activity) for each DOI. 
The intuition is that, because DOI attention typically shows a bursty, log-normal pattern, this interval captures the main triggering event while excluding early noise and long trailing mentions.
% The intuition is that DOI activity typically follows a log-normal distribution, with a burst of attention at the start of high-traction events following long tails with more sparse mentions.
% Thus, this region should capture the ``triggering'' event that drives attention while excluding sporadic mentions before, and long decay tails after, this event.
We visually show this 50\% density region distribution for the top 4 most-mentioned papers in the Appendix. 
To ensure robustness, we also perform backward lookups of 3, 6, and 12 hours before the detected start date and take the earliest instance within these periods as the start of the high-density region. 
This sensitivity analysis, reported in the Appendix, produces nearly identical results.

Within this high-density region, we identify the density peaks for conformist (pro-consensus) superspreaders (PSPs), contrarian (anti-consensus) superspreaders (ASPs), low-credibility (LC), and high-credibility (HC) outlets.
Outlets are binarized into LC and HC using a trust score cutoff computed as Youden's J (see Appendix) for interpretability. 
For each DOI, we determine the individual density peak of each class, and derive whether the peak of one class precedes the other.

\begin{figure}[t!]
    \includegraphics[width=0.99\linewidth]{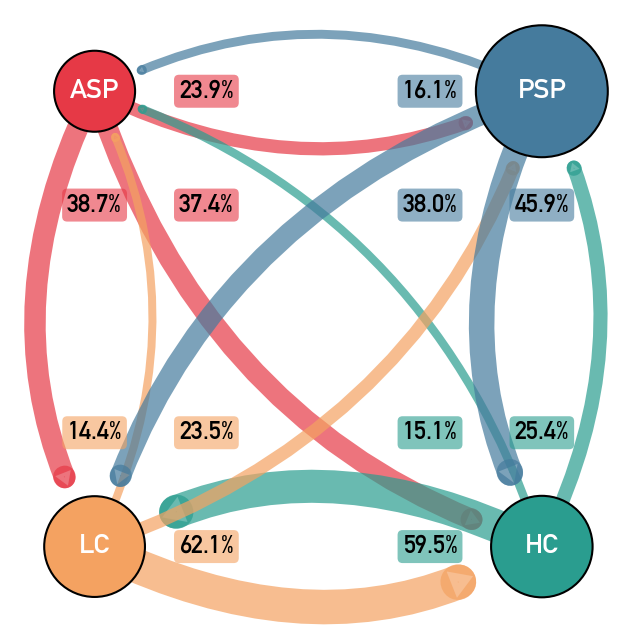}
    \caption{Precedence pathways.}
    \label{fig:sp_media_paths}
\end{figure}

% retain only the first instance of each of the four classes. 
% This is based on the intuition that the first appearance of a class triggers further activity from that class due to the log-normal distribution, making the first instance particularly informative. 
We then construct a directed network showing this precedence across all DOIs among the classes. 
We draw pairwise edges between classes and, as we iterate through DOIs, increment the edge weight whenever one class's density peak precedes another's.
For example, if ASP's peak appears before LC, we increment the ASP $\rightarrow$ LC edge weight by one. 
We scale the edge weights row-wise so that each class's outgoing edges sum to 100 (i.e., percentages). 
These relationships are visualized in Figure~\ref{fig:sp_media_paths}.
% presented separately between the two types of superspreaders for clarity.
% The full 4-class pathways and robustness checks are provided in the Appendix. 

These information pathways reveal that superspreaders typically precede news outlets.
Taking into account base posting frequencies, contrarian superspreaders are significantly more likely to precede both low-quality ($\chi^2_{(1)} = 127.56, p < 0.001$, median $\Delta_{time} = 28h$) and high-quality ($\chi^2_{(1)} = 123.41, p < 0.001$, median $\Delta_{time} = 23h$) news.
Similarly, conformist superspreaders are more likely to precede both high-quality ($\chi^2_{(1)} = 235.33, p < 0.001$, median $\Delta_{time} = 23.5h$) and low-quality ($\chi^2_{(1)} = 166.08, p < 0.001$, median $\Delta_{time} = 19.6h$) news.
On the contrary, there are no significant differences in the precedence order of low- and high-quality news ($\chi^2_{(1)} = 1.87, p = 0.17$) or contrarian and conformist superspreaders ($\chi^2_{(1)} = 0.51, p = 0.47$), indicating a more generalized Twitter-to-news information pathway.
However, we stress that this analysis does not establish any causal information flows.
Rather, it demonstrates that scientific papers typically receive traction on Twitter before being picked up by news media, whose articles may subsequently be re-introduced to a wider platform base~\cite{alperin_second-order_2024,west_misinformation_2021}.

\subsection{Summary (RQ2)}

Overall, we find overlap in the science dissemination of superspreaders and news media: while conformist superspreaders discuss papers picked up mostly by mainstream and higher-trust media, contrarian superspreaders align more with low-quality media like conspiratorial or pseudoscientific outlets in the science they share.
Moreover, although we cannot claim causality, we find that, on average, science tends to be discussed by superspreaders on Twitter first before being picked up in news media.

\section{Discussion}

In this paper, we examine the roles of different actors involved in the communication of COVID-19 science.
We make several important observations, including how the obfuscation of scientific consensus~\cite{beers_selective_2023,efstratiou_heres_2024} and centralization of perceived experts~\cite{harris_perceived_2024} noted in other works may arise.
We find that coordinated accounts amplify targeted experts who possess the credentials to introduce (selective) science into their wider networks.
Although the majority of influencers driving scientific discussions on Twitter are non-contrarian science communicators, coordination seems to be strongest around those contrarian superspreaders.

Moreover, we find that news media mirror the roles of superspreaders.
While mainstream media tend to report on science that is also shared by pro-consensus experts, lower-trust outlets are aligned with the science reporting activity of contrarian superspreaders in terms of the papers they cover.
Our findings paint a comprehensive picture of how science communication and its aberrant forms take place across different media and entities of various influence and roles.

Furthermore, we add to the literature on the interplay of news media and platforms.
While news articles about science may garner more attention on social media than scientific articles themselves~\cite{alperin_second-order_2024}, we show that, in addition to social media conversations following these articles~\cite{smeros2019scilens,wright_modeling_2022}, news media reporting on specific scientific papers often follows discussions of these papers by influential figures on Twitter.
Our findings indicate possible ``closed loops'', where platform influence may precede news reporting, with this news reporting then driving further platform attention.

This work highlights the multi-level landscape of online science communication, both within the same platform and between media.
We show that seemingly disjointed activity may compound, as in the case of coordinated networks amplifying selected experts, or the boosting of selective pieces of science by high-influence actors like superspreaders and news outlets with substantial followings.
Although multi-platform information diffusion is already a broad area of interest for researchers~\cite{gerard_bridging_2025,wilson_cross-platform_2021}, we demonstrate the importance of these considerations to science communication as well.
At a time when many scientific topics are polarized and draw in actors such as government agencies, political commentators, and other non-scientific influencers, it is essential to understand how these groups collectively shape science communication, whether their participation is genuine or opportunistic.
% Particularly at a time where several scientific topics are polarized or politicized (and thus involve more actors like government agencies, political commentators, or other non science-related influencers), the compounding effects of these different entities partaking in science communication as a collaborative exercise involving genuine and potentially malicious actors alike is crucial to understand.

However, there are some limitations associated with our work.
First, we only consider news and tweets about scientific \textit{papers} specifically.
Although this is indeed the intended scope, there may be other avenues through which science communication is conducted.
Second, these information pathways concern a specific time period and topic (COVID-19) on a specific platform (Twitter).
Therefore, the generalizability of our findings to other platforms or scientific topics is limited.
Finally, especially with respect to media and Twitter relationships, the patterns we report cannot be deemed to be causal, but rather as descriptive accounts of the types of events that tend to precede one another.

\section{Acknowledgments}

This work has been supported by the University of Washington’s Center for an Informed Public, the John S. and James L. Knight Foundation (G-2019-58788), and the William and Flora Hewlett Foundation (2023-02789).

\small
\bibliography{cleaned_references,extended}

\normalsize

% \clearpage
% \input{ethics}

\clearpage
\appendix
\section*{Appendix}

\subsection{Coordination Cutoff Robustness Checks}

\begin{figure*}[t!]
    \centering
    \begin{subfigure}[t]{0.33\linewidth}
        \centering
        \includegraphics[width=0.99\linewidth]{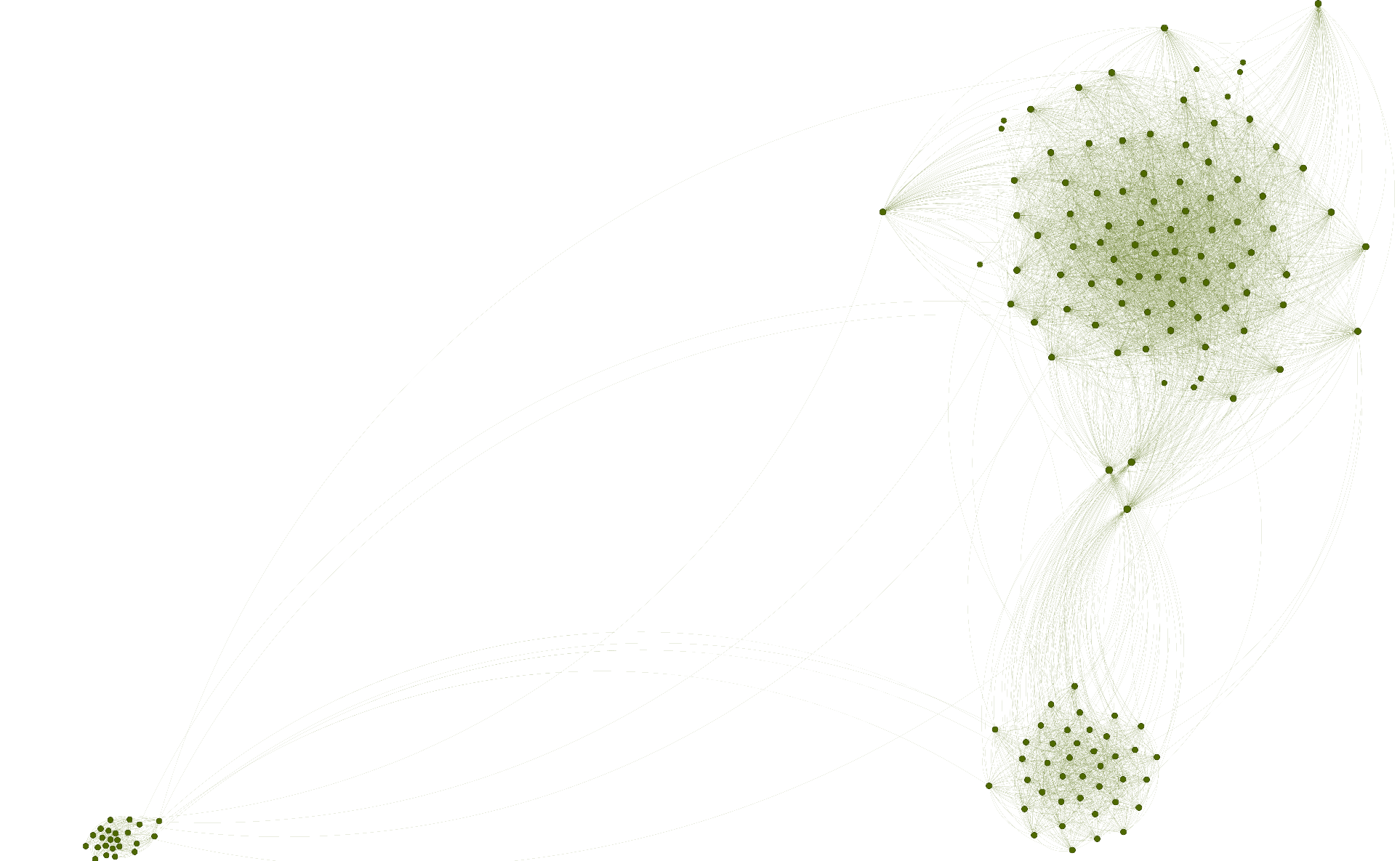}
        \caption{0.5\% threshold.}
        \label{subfig:coord_05}
    \end{subfigure}
    \begin{subfigure}[t]{0.33\linewidth}
        \centering
        \includegraphics[width=0.99\linewidth]{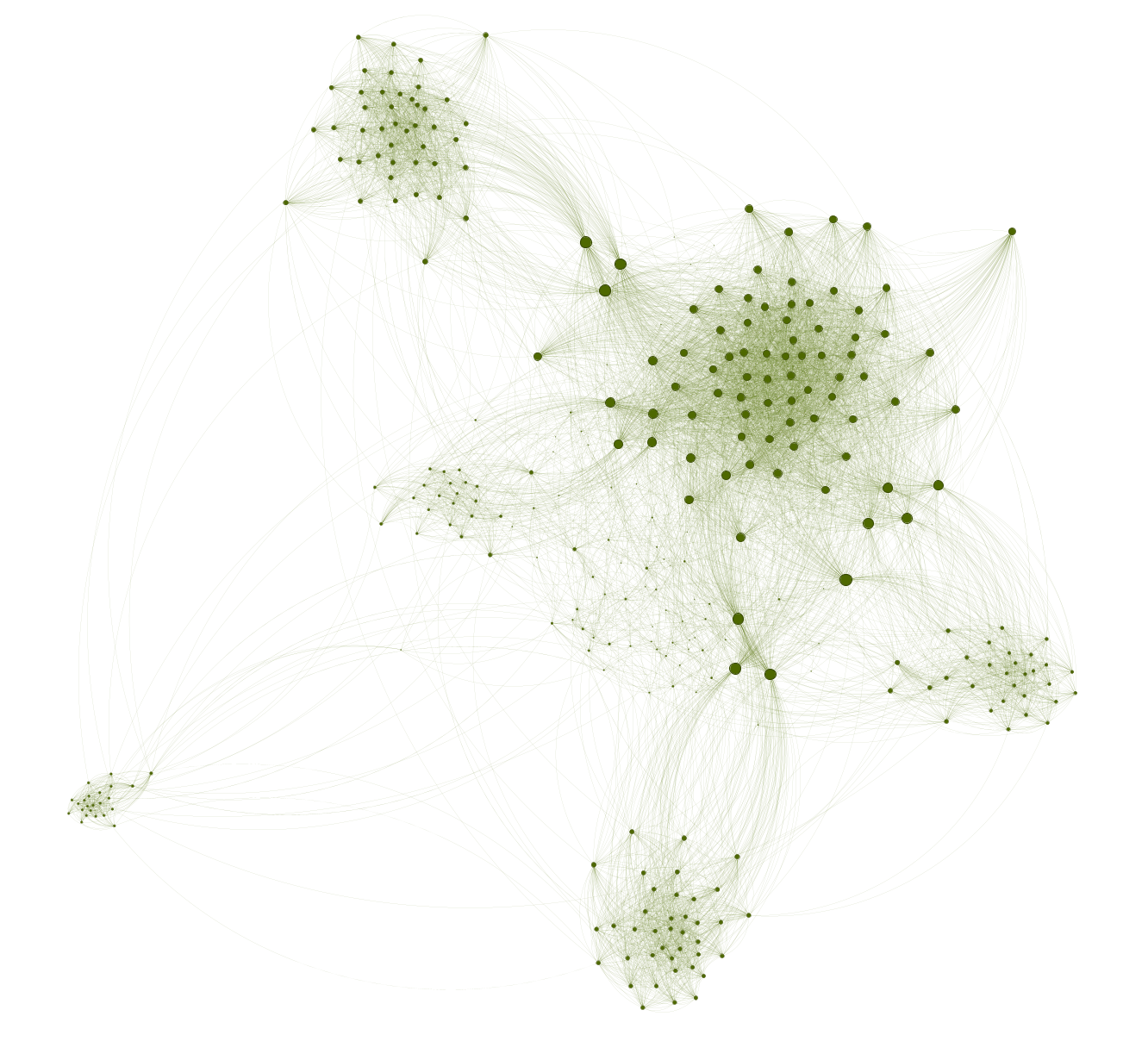}
        \caption{1\% threshold.}
        \label{subfig:coord_1}
    \end{subfigure}
    \begin{subfigure}[t]{0.33\linewidth}
        \centering
        \includegraphics[width=0.99\linewidth]{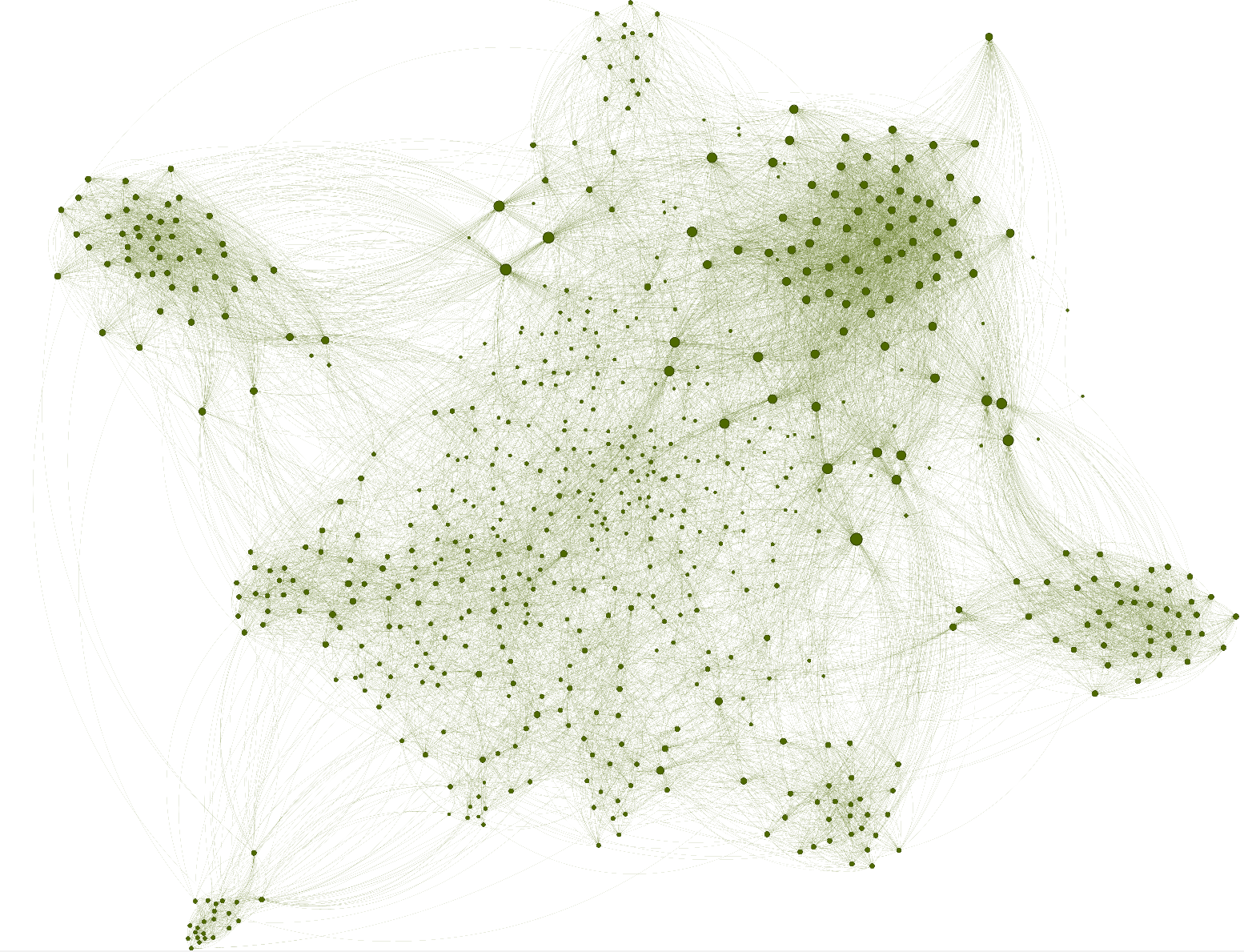}
        \caption{2\% threshold.}
        \label{subfig:coord_2}
    \end{subfigure}
    \caption{Visual coordinated network structures with different centrality thresholds.}
    \label{fig:coord_compare}
\end{figure*}

To ensure that the network structure is not an artifact of our eigenvector centrality threshold, we compare our 1\% top-centrality coordinated network to thresholds of 0.5\% and 2\%.
Specifically, we examine visual network structure (Figure~\ref{fig:coord_compare}), as well as statistics like the percentage of contrarians under each configuration, number of nodes, density, etc. (Table~\ref{tab:coord_stats}).

\begin{table}[t]
    \centering
    \begin{tabular}{lrrr}
    \toprule
        \textbf{Statistic} & \textbf{0.5\%} & \textbf{1\%} & \textbf{2\%} \\
        \midrule
        \textbf{\#nodes} & 153 & 306 & 612 \\
        \textbf{\%contrarian} & 95.4 & 96.4 & 96.7 \\
        \textbf{Density} & 0.379 & 0.15 & 0.062 \\
        \textbf{Modularity} & 0.5 & 0.617 & 0.614 \\
        \textbf{Avg. path length} & 1.84 & 2.16 & 2.37 \\
        \textbf{Avg. clustering coefficient} & 0.957 & 0.814 & 0.675 \\
        \bottomrule
    \end{tabular}
    \caption{Network statistics by coordination centrality threshold.}
    \label{tab:coord_stats}
\end{table}

The patterns we observe show that the coordination detection method is robust to the centrality threshold, as the coordinated network simply expands in number of nodes instead of resulting in a different network. 
In every case, the networks have a diameter of 4 and show a single connected component.
The percentage of contrarians remains high (above 95\%) for all three thresholds.
% As is expected, there is a linear increase in the number of nodes and average path length, and a linear decrease in the density and average clustering coefficient as the threshold increases; in other words, the changes in these statistics are (unsurprisingly) commensurate with the change in the centrality threshold.
We also observe a nearly identical modularity for the thresholds of 1\% and 2\%, which is higher than the modularity for a threshold of 0.5\%.
Looking at Figure~\ref{fig:coord_compare}, this is because the higher thresholds capture more of the clusters that are adjacent to the coordinated center.
Beyond this, however, the differences between the different thresholds are not structural, but rather, completeness-related (i.e., higher thresholds naturally capture more nodes and more edges).

\subsection{Manual Annotation of Superspreaders}

The manual assessment of superspreader accounts was conducted by one of the authors who is closely familiar with scientific discourse of COVID-19 on Twitter, and prominent scientific voices in the discourse.
The profile of each account (handle, bio, number of followers etc.) was manually assessed to determine whether it fell into any of the following categories:

\begin{enumerate}
    \item \textbf{Medical doctor} (e.g., MDs, physicians, etc.) For this classification, we look for evidence of a person possessing medical knowledge that enables them to practice medicine on patients.
    \item \textbf{Scientist} (e.g., researcher, academic, etc.) We restrict this classification to people working in COVID-adjacent fields, e.g., medicine, biology, virology, epidemiology, etc. We do not classify other types of scientists as such if they work in unrelated fields (e.g., physics, computer science, social sciences, etc.), unless they work in those fields in a COVID-related capacity (e.g., physical virology, computational epidemiology, public health, etc.) For this label, we only consider people who worked in their respective fields in an \textit{official} capacity, e.g., a public health role, academic or industry research role, etc., unless they have otherwise demonstrable expertise (e.g., professional society memberships, recognized accreditations, published peer-reviewed research on COVID-19, etc.)
    \item \textbf{Scientific organization} (e.g., science-related agencies, scientific journals, scientific or medical companies, science NGOs, academic departments or labs). Note that this does not include non-scientific government agencies or other organizations not related to science.
    \item \textbf{Science communicator} (e.g., high-profile science personalities, science journalists, scientific news aggregators, etc.) This does not include people who do not explicitly or predominantly focus on science for their content or reporting (e.g., journalists who only covered science during the pandemic).
    \item \textbf{Other scientific authority}. This includes any other science-related accounts that do not fall in any of the above categories.
    \item \textbf{Non-scientific}. Any account that cannot be classified into any of the above. 
\end{enumerate}

Whenever the annotator encountered accounts that they were unfamiliar with, they verified the account's identity through the account owner's online presence before making an annotation decision (e.g., an existing physician practice for MDs, Google Scholar profiles for researchers, other relevant digital footprints like Wikipedia pages or authored books, etc.)
We assumed negligible risk of impersonation, as impersonation profiles would have been unlikely to reach the prominence of these superspreader accounts without being detected and taken down.
If a determination could not be made, or if an account could not be verified (e.g., due to the absence of real, traceable names), a label of ``non-scientific'' was applied.

In several cases, accounts could be classified in more than one category (e.g., medical doctors doing active research).
In cases where a person held a medical doctorate (MD) or had previously worked as a physician, a medical doctor label took precedence to indicate medical knowledge even if their primary function was that of a researcher (i.e., ``medical doctor'' was taken in a literal sense).
In other cases (e.g., where someone was both a scientist and science communicator), the label was based on the account's primary function and/or what the account was primarily known for, as determined through bio self-identification, personal or Wikipedia pages, Google profiles, etc.

% examples of none: unverifiable accounts (i.e., undisclosed names etc.), investors/biotech industry non-science ppl, COVID survivors/health activists

\subsection{Extended Topical Distributions}

\begin{figure*}[t!]
    \centering
    \begin{subfigure}[t]{0.33\linewidth}
        \centering
        \includegraphics[width=0.99\linewidth]{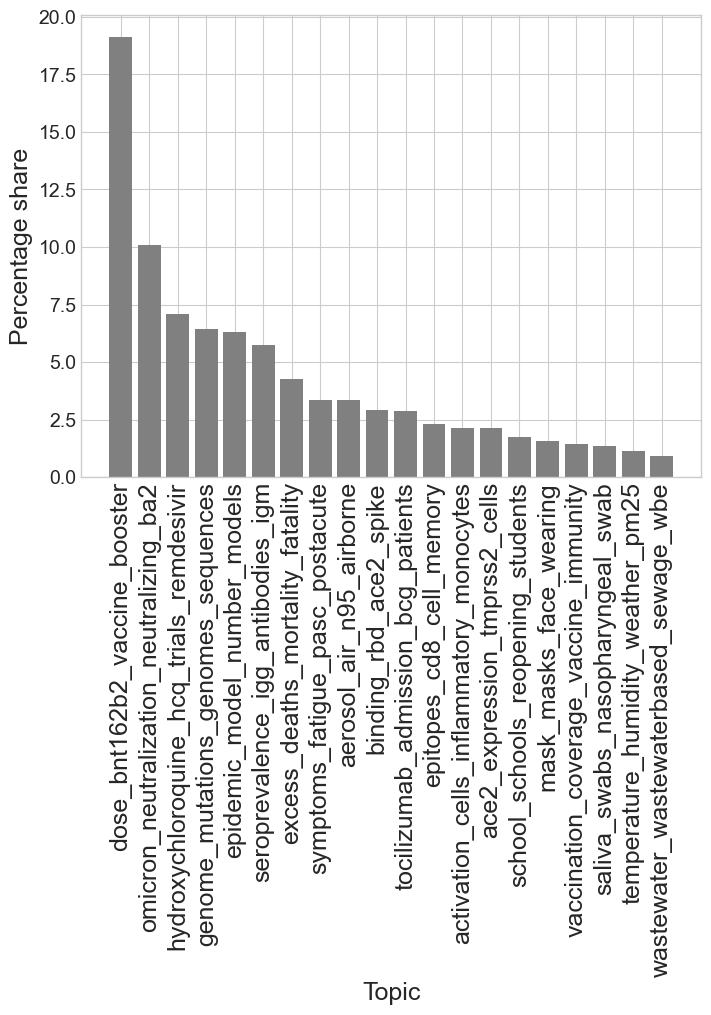}
        \caption{Overall distribution.}
        \label{subfig:share_total}
    \end{subfigure}
    \begin{subfigure}[t]{0.33\linewidth}
        \centering
        \includegraphics[width=0.99\linewidth]{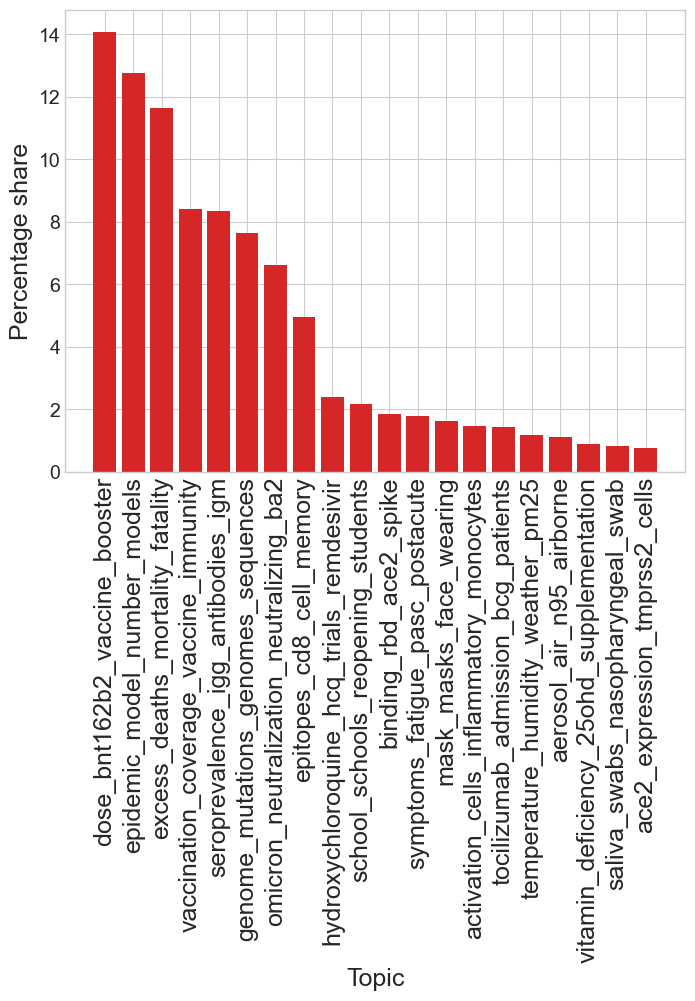}
        \caption{Contrarian superspreaders.}
        \label{subfig:share_anti}
    \end{subfigure}
    \begin{subfigure}[t]{0.33\linewidth}
        \centering
        \includegraphics[width=0.99\linewidth]{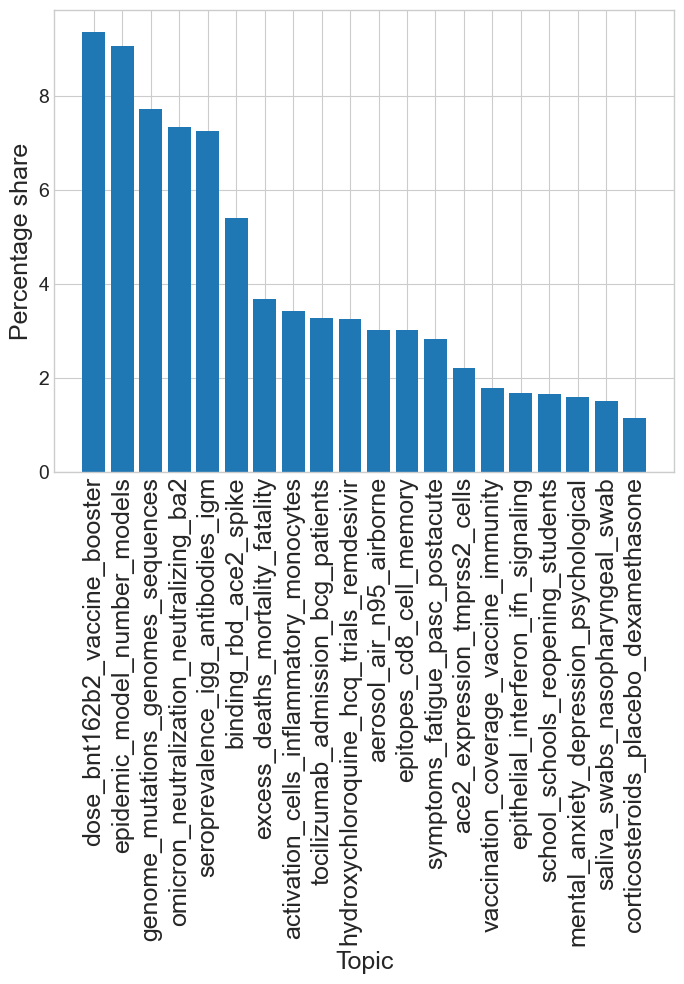}
        \caption{Conformist superspreaders.}
        \label{subfig:share_pro}
    \end{subfigure}

    \caption{Distribution of the top 20 topics by group. X-axis labels are the most representative topical terms.}
    \label{fig:topicshares}
\end{figure*}

Here, we show the distribution of topical shares for contrarian superspreaders, conformist superspreaders, and the overall network. 
The topic model detected 53 topics overall (excluding the outlier cluster). 
Due to some topics having negligible prevalence ($>$ 1\%) and for better visualization, we plot the top 20 for each group in Figure~\ref{fig:topicshares}.

\subsection{Syndication Analysis}

\begin{figure}[h]
    \centering
    \includegraphics[width=0.99\linewidth]{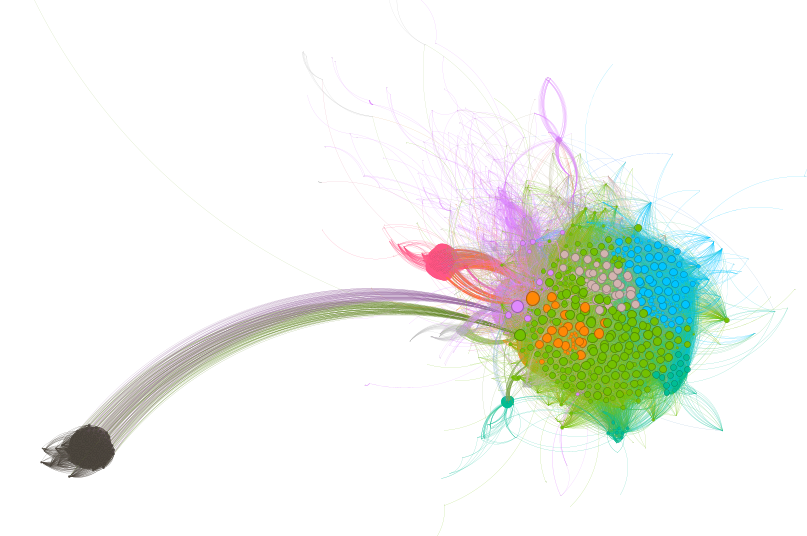}
    \caption{Syndicated outlet neighborhoods.}
    \label{fig:syndication}
\end{figure}

We perform a syndication analysis to ensure that any media coordination is not anomalous.
We take outlets that shared the exact same title within an hour of each other and form undirected edges between them; the resulting network is shown in~ Figure~\ref{fig:syndication}.
Typically, neighboring outlets are those under the same brand (e.g., NBC-affiliated outlets) or local networks coordinating to share national news to local communities.
The deviating black cluster represents an Australian high-quality news neighborhood; therefore, the deviation is due to locality and does not suggest suspicious activity.

\subsection{Distribution of Paper Mentions Over Time}

\begin{figure*}
    \includegraphics[width=0.99\linewidth]{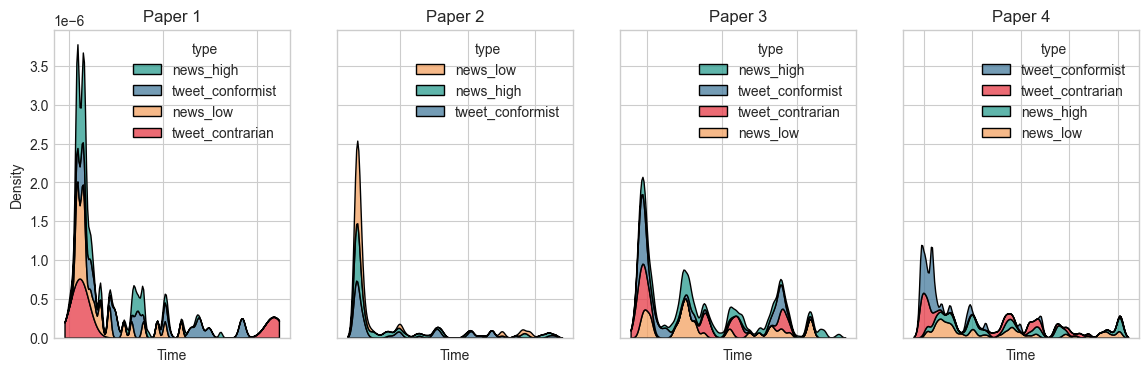}
    \caption{Over-time density distributions for most mentioned papers.}
    \label{fig:top4_distros}
\end{figure*}

Figure~\ref{fig:top4_distros} shows the density of mentions over time for the top 4 most-mentioned papers in the dataset, within the 50\% density region.
As can be seen, papers typically receive a burst of attention, followed by longer tails of more sporadic activity.

\subsection{Robustness Checks for Precedence Pathways}

\begin{figure*}[t!]
    \centering
    % \begin{subfigure}[t]{0.49\linewidth}
    %     \centering
    %     \includegraphics[width=0.99\linewidth]{figures/pathways.png}
    %     \caption{No lookup.}
    %     \label{subfig:bl_nolookup}
    % \end{subfigure}
    \begin{subfigure}[t]{0.33\linewidth}
        \centering
        \includegraphics[width=0.99\linewidth]{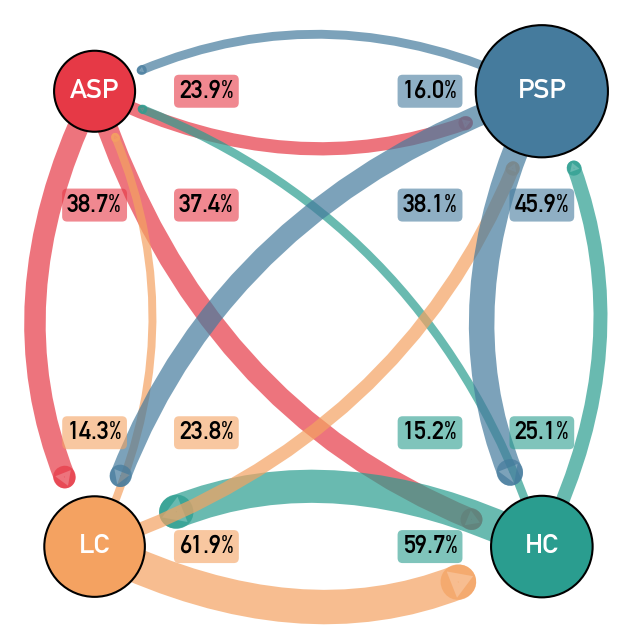}
        \caption{3-hour lookup.}
        \label{subfig:bl_3h}
    \end{subfigure}
    \begin{subfigure}[t]{0.33\linewidth}
        \centering
        \includegraphics[width=0.99\linewidth]{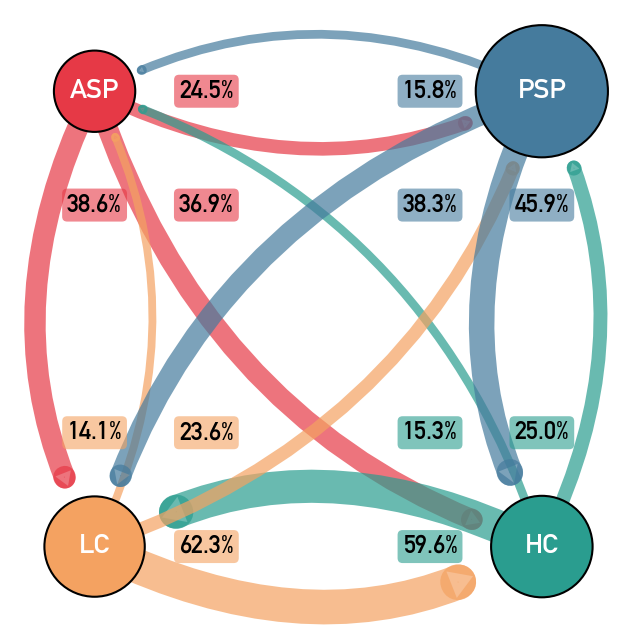}
        \caption{6-hour lookup.}
        \label{subfig:bl_6h}
    \end{subfigure}
    \begin{subfigure}[t]{0.33\linewidth}
        \centering
        \includegraphics[width=0.99\linewidth]{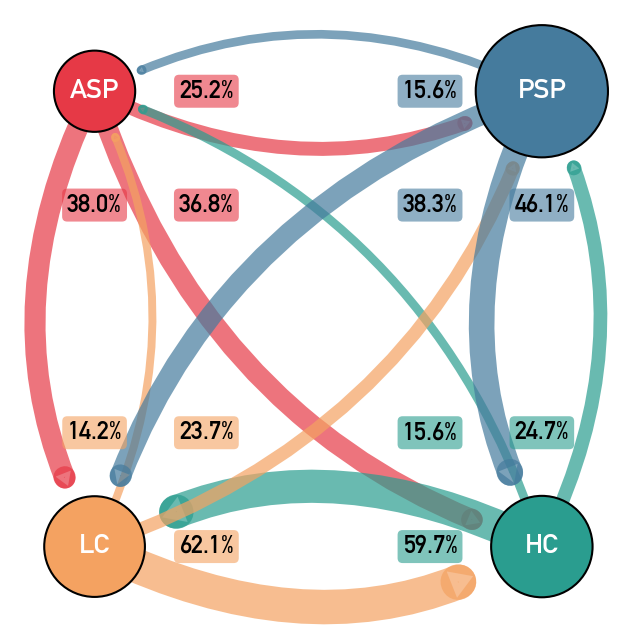}
        \caption{12-hour lookup.}
        \label{subfig:bl_12h}
    \end{subfigure}
    \caption{Backward lookup robustness checks.}
    \label{fig:robustness}
\end{figure*}

In Figure~\ref{fig:robustness}, we show robustness checks for the media-superspreader precedence links by range of the backward lookup.

\subsection{Outlet Binarization}

The binarization into LC and HC outlets is done using a cutoff in the 11.5k outlet trust scores reported by~\citet{lin_high_2023}.
To derive this cutoff, we use labeled datasets from~\citet{baly_predicting_2018} and~\citet{baly_what_2020}, who use ``low'', ``mixed'', or ``high'' factuality labels based on Media Bias/Fact Check (MBFC) ratings of 1.42k unique domains.
We collapse low- and mixed-factuality sources into the ``questionable'' class, and use high-factuality sources as ground-truth for the ``trustworthy'' class.
We treat these scores (ranging from 0 to 1) as prediction probability estimates and obtain the ROC curve, from which we compute Youden's J (i.e., the optimal point at which the true positive minus the false positive rate is maximized).
We find that \textit{J} = 0.647, which we use as our cutoff point (F1 score = 0.83).

\end{document}